\documentclass[final,3p,times]{elsarticle}
\usepackage{graphicx, float, subcaption, enumitem}
\usepackage{amsfonts, hyperref, siunitx, multicol}
\usepackage{amsmath, mathtools, cancel, nicefrac, multirow}
\usepackage[ruled,vlined]{algorithm2e}
\usepackage[dvipsnames]{xcolor}
\usepackage{booktabs, appendix}

\newcounter{bla}

\journal{Computer Physics Communications}

\begin{document}
\begin{frontmatter}

\title{HADOKEN: An Open-Source Software Package for Predicting Electron Confinement Effects in Various Nanowire Geometries and Configurations}
\author{Bryan M. Wong*}
\author{Cameron Chevalier}

\cortext[author] {Corresponding author.\\\textit{E-mail address:} bryan.wong@ucr.edu, \textit{Webpage:} http://www.bmwong-group.com}
\address{Department of Chemical \& Environment Engineering, Materials Science \& Engineering Program, Department of Physics \& Astronomy, and Department of Chemistry\\University of California-Riverside, Riverside, California 92521, United States}

\begin{figure}
  \centering
  \includegraphics[width=150mm]{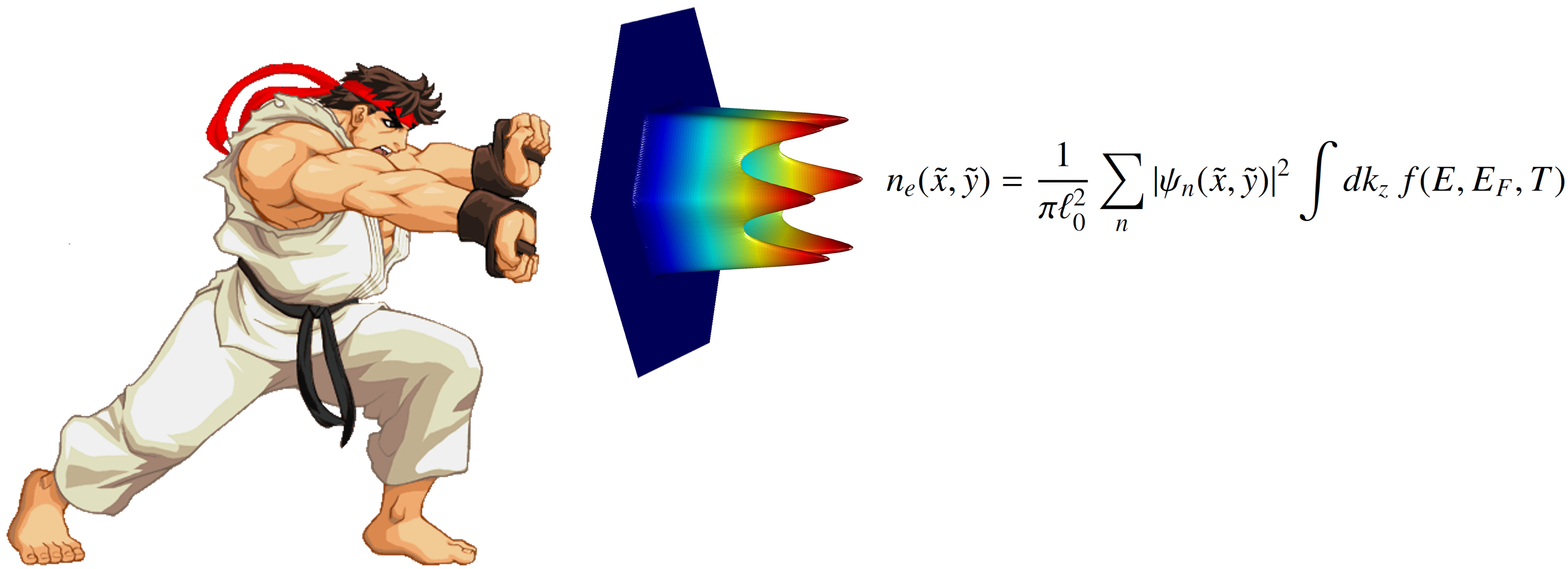}
  \caption*{ }
\end{figure}

\begin{abstract}
We present an open-source software package, HADOKEN (High-level Algorithms to Design, Optimize, and Keep Electrons in Nanowires), for predicting electron confinement/localization effects in nanowires with various geometries, {arbitrary number of concentric shell layers,} doping densities, and external boundary conditions. The HADOKEN code is written in the MATLAB  programming environments to aid in its readability and general accessibility to both users and practitioners. We provide several examples and outputs on a variety of different nanowire geometries, boundary conditions, and doping densities to demonstrate the capabilities of the HADOKEN software package. As such, the use of this predictive and versatile tool by both experimentalists and theorists could lead to further advances in both understanding and tailoring electron confinement effects in these nanosystems.
\end{abstract}


\end{frontmatter}
{\bf PROGRAM SUMMARY}

\begin{small}
\noindent
{\em Program Title:} HADOKEN\\
{\em CPC Library link to program files:} \verb"https://doi.org/10.17632/jyzk4gfytx.1" \\
{\em Licensing provisions:} GNU General Public License 3 \\
{\em Programming language:} MATLAB\\
{\em Nature of problem:} HADOKEN utilizes iterative finite element methods to solve coupled Schr\"odinger and Poisson equations for heterostructure core--shell nanowires with arbitrary cross-sectional geometries. The user-friendly program outputs graphical results of electronic energies, densities, wavefunctions, and band profiles for various user-supplied input parameters.\\
{\em Solution method:} iterative solution of coupled Schr\"odinger and Poisson equations using finite element methods and sparse matrix linear algebra.\\
\\
\end{small}

\section{Introduction}\label{intro}
Semiconductor nanowires (NWs) continue to garner significant interest in various applications ranging from next-generation electronics to nanoscale probes for biological systems \cite{hM09,pY10,eG19,mC14}. With cross-sectional dimensions tailorable to a few nanometers, these systems allow quantum confinement effects to emerge as electrons become quantized into discrete energy levels \cite{dT82,gG20,kP15}. In particular, core--shell nanowires give rise to additional quantum effects since mobile two-dimensional electron gases (2DEGs) can form at the semiconductor--semiconductor heterojunction interface \cite{mG02,mB02,wL05,fQ04,mf13,cL19}. To fully harness the electronic properties of these systems, a wide range of material properties (such as doping density, bandgap alignment, geometry, and structural composition) may be altered to achieve spontaneous electron gas formation \cite{lL02,sF13,aS19}. While the resulting parameter space is immense, theory and predictive modeling provide a guided path for determining which combination of material properties/parameters best optimizes performance of these novel nanosystems.

This work presents an open-source software package, HADOKEN (High-level Algorithms to Design, Optimize, and Keep Electrons in Nanowires), for predicting the formation of electron gases in core--shell nanowires with arbitrary geometries/shell layers, doping densities, and external boundary conditions. The code utilizes a self-consistent numerical implementation that solves coupled Schr\"odinger and Poisson equations to obtain wavefunctions, electron densities, and band-bending diagrams \cite{lW06,aB11}. HADOKEN is written in the MATLAB programming environment to aid in its readability and general accessibility to both users and practitioners. Since open-source Schr\"odinger--Poisson codes for arbitrary core--shell geometries and boundary conditions are not readily accessible, our publicly available HADOKEN code provides a user-friendly program for researchers by reducing the time commitment of writing these complex algorithms from scratch. To demonstrate its utility, we extensively document and provide several examples of different nanowire configurations that can be handled by the HADOKEN software package.

This paper is structured as follows: Section~\ref{theory} introduces the physical systems considered in our calculations and the governing equations that are solved numerically. Section~\ref{methods} provides additional implementation details for each of the algorithms used in HADOKEN. Section~\ref{results} presents typical results for a variety of nanowire geometries, configurations, and boundary conditions. The outputs for each computed system are also analyzed and given a physical justification. Section~\ref{conclusions} then concludes with a summary and future perspective on various potential applications of the HADOKEN program.

\section{Theory and Methodology}\label{theory}
\begin{figure}[t]
	\centering
	\begin{subfigure}[t]{0.495\textwidth}
		\centering
		\includegraphics[width=47.099mm]{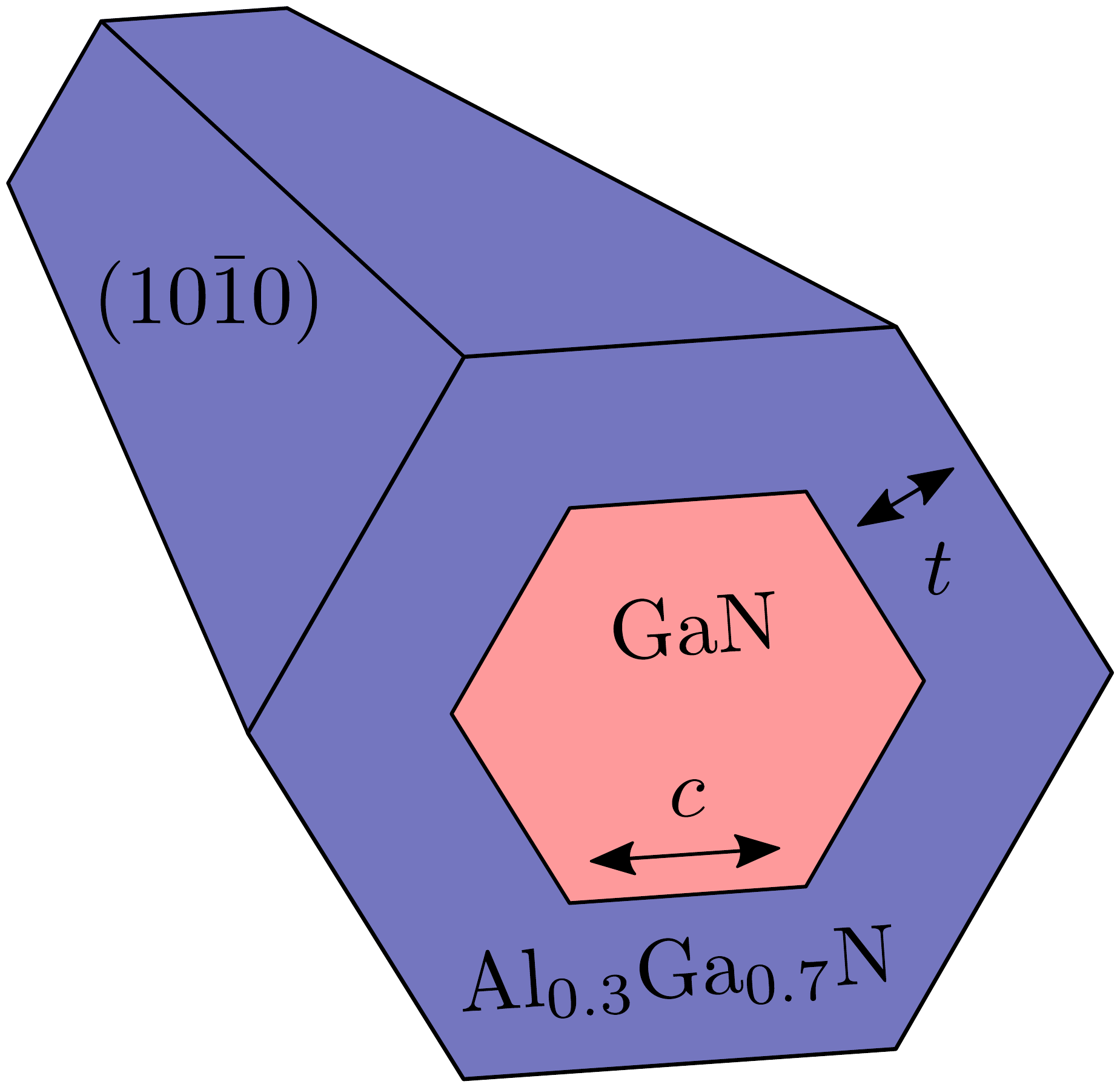}
		\caption{}
		\label{F:hexCross}
	\end{subfigure}
	\hfill
	\begin{subfigure}[t]{0.495\textwidth}
		\centering
		\includegraphics[width=60.843mm]{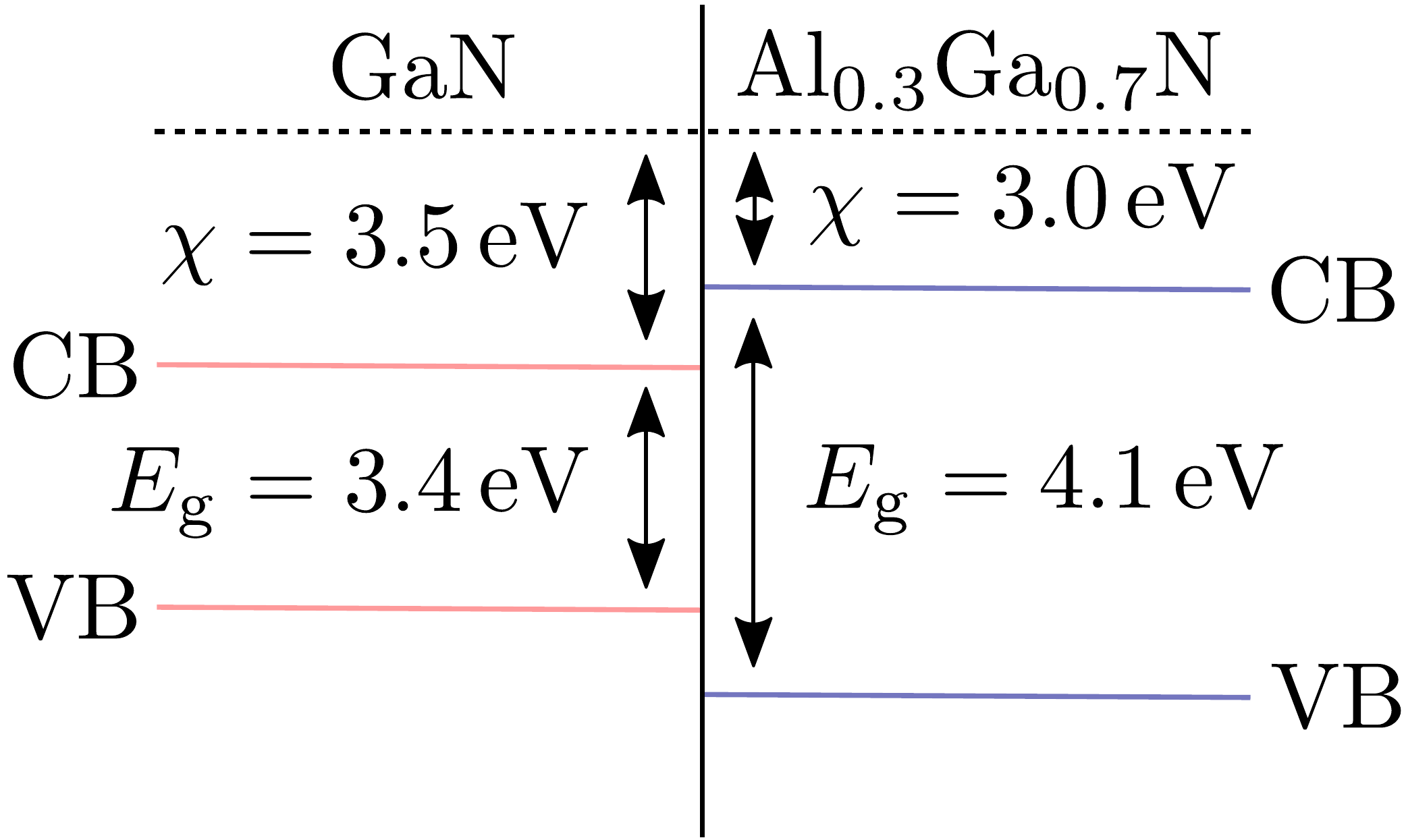}
		\caption{}
		\label{F:bandDiagram}
	\end{subfigure}
	\hfill
	\begin{subfigure}[t]{0.495\textwidth}
		\centering
		\includegraphics[width=43.970mm]{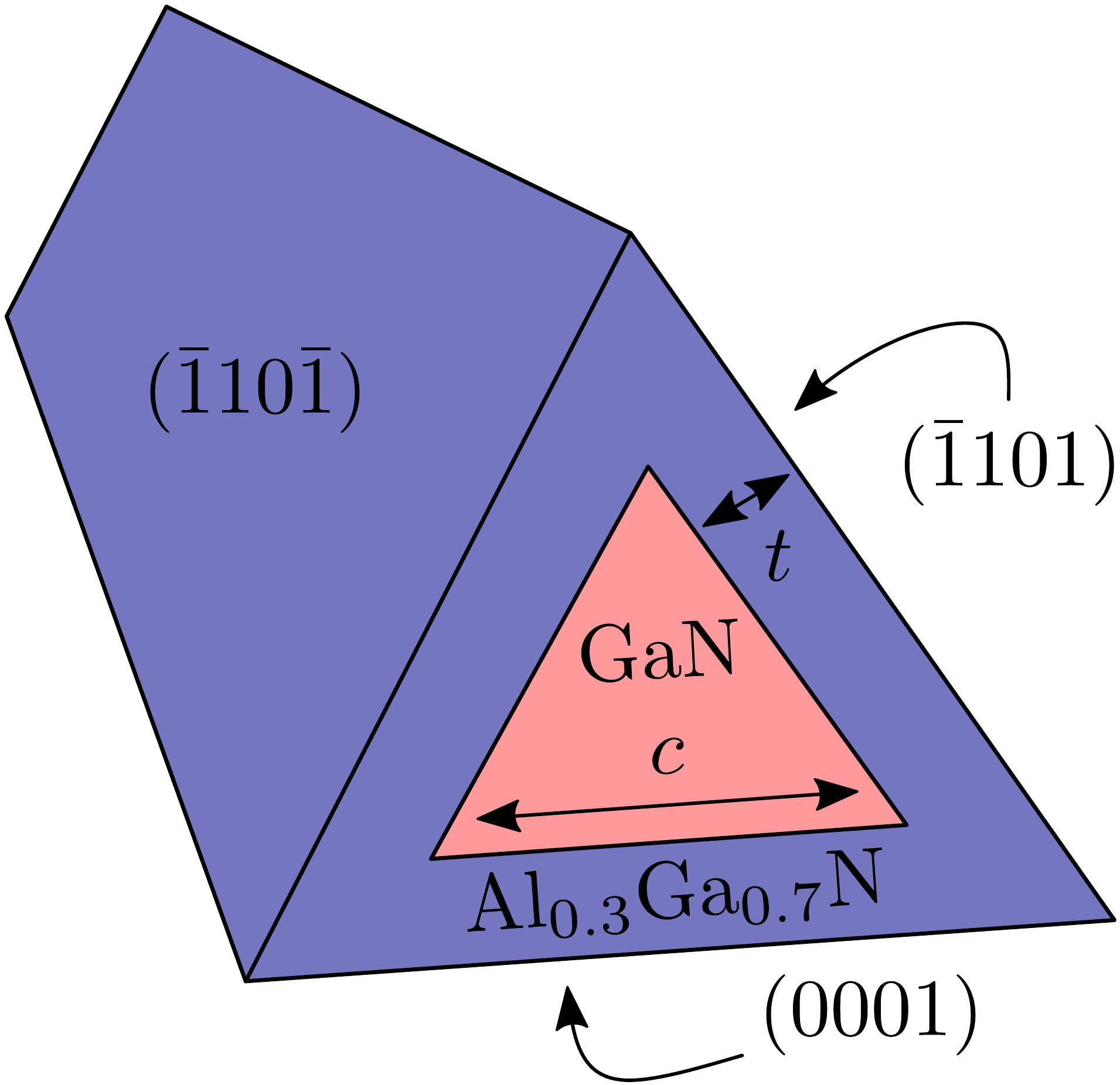}
		\caption{}
		\label{F:triCross}
	\end{subfigure}
	\hfill
	\begin{subfigure}[t]{0.495\textwidth}
		\centering
		\includegraphics[width=79.194mm]{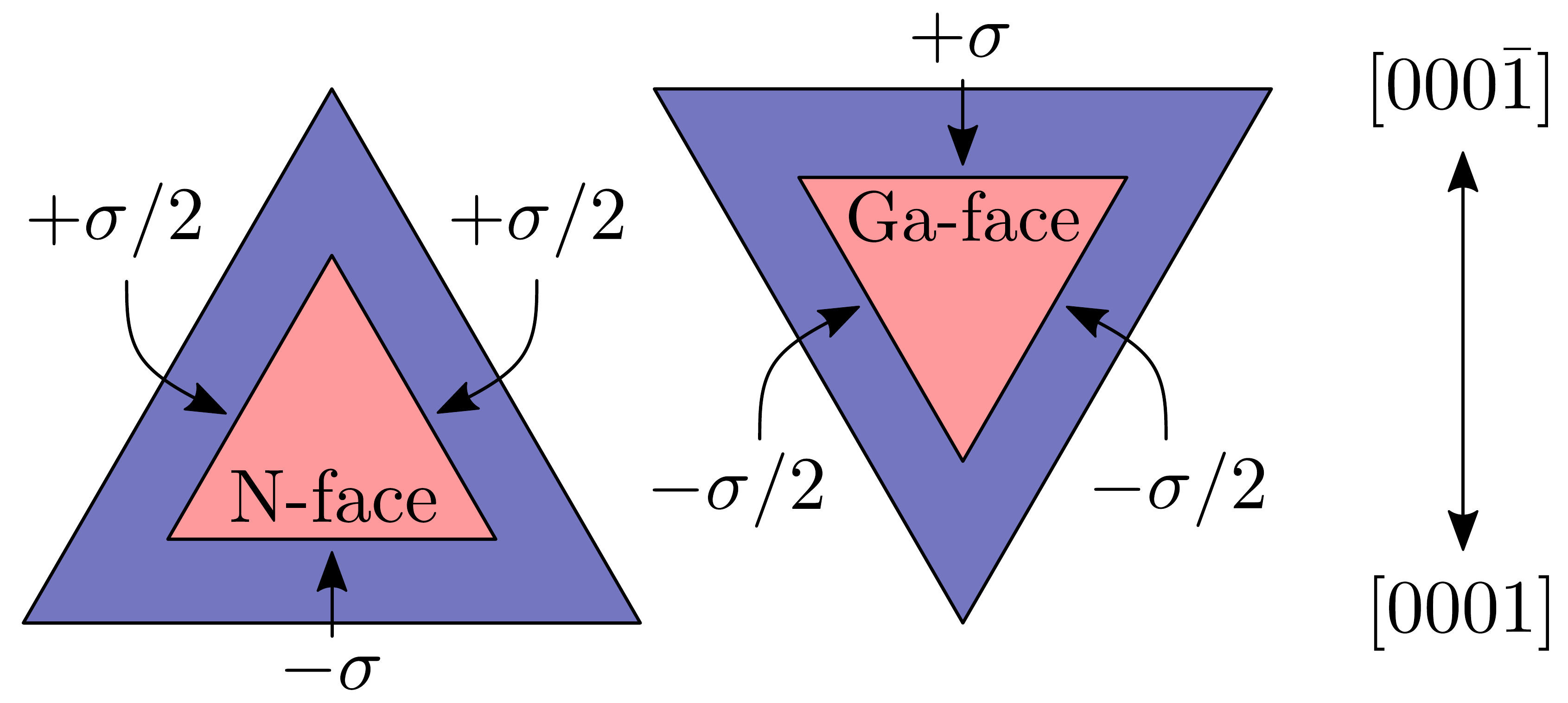}
		\caption{}
		\label{F:triPolars}
	\end{subfigure}
	\caption{Schematics of the (a)\,hexagonal and (c)\,triangular GaN/AlGaN core--shell NWs considered in this work. (b)\,Valence band (VB) and conduction band (CB) alignment at the core--shell interface. The $\Delta E_c = 0.5\,\text{eV}$ discontinuity between the conduction bands of each material establishes a two-dimensional quantum well in the  cross-sectional plane of the NW. (d)\,Two possible crystallographic orientations of the triangular heterostructure: the $(0001)$ Ga-face and $(000\bar{1})$ N-face orientations. Each orientation has one polar interface with a charge density of $\sigma=\pm 0.0156$ C/m$^2$ and two semi-polar interfaces with a charge density of $\sigma/2= \mp 0.0078$ C/m$^2$}
	\label{F:schematic}
\end{figure}

Fig.~\ref{F:schematic} depicts the NW examples considered in this work, which have either hexagonal or triangular cross-sections (the latter has two different crystallographic orientations). {While Fig.~\ref{F:schematic} depicts a single GaN/AlGaN core--shell configuration for simplicity, HADOKEN can calculate electronic properties for core--multishell NWs with arbitrary cross-sections, numbers of concentric layers, and material compositions as well.} We feature the hexagonal and triangular NW cross-sections in this work since these geometries/orientations have been experimentally observed and synthesized \cite{fQ04,fQ05,yL06,kH10,tK04,gW06,nS05,oH05}. For the single core--shell configuration depicted in Fig.~\ref{F:schematic}, each NW is composed of an $\text{Al}_{0.3}\text{Ga}_{0.7}\text{N}$ shell with uniform thickness, $t$, encompassing a GaN core of side length $c$ (from simple geometry, the shell side length, $s$, is related to $t$ and $c$ via the expressions $s = 2t / \sqrt{3} + c$ and $s = 2 \sqrt{3} t + c$ for hexagonal and triangular cross-sections, respectively). For the hexagonal NW, the axis is aligned in the $[0001]$-direction, and the cross-section is bounded by $\{10\bar{1}0\}$ planes. For each of the triangular NWs, the axis is aligned in the $[11\bar{2}0]$-direction, and the cross-section is bounded by two equivalent $(\bar{1}10\bar{1})$ and $(\bar{1}101)$ planes, and a $(0001)$ plane. As shown in Fig.~\ref{F:triPolars}, the triangular NWs have two possible orientations of the $(0001)$ plane---either in the $[000\bar{1}]$- or $[0001]$-direction---which correspond to physically distinct configurations. In the scientific literature, the former is referred to as an N-terminated face and the latter a Ga-terminated face.

The electronic parameters of the GaN core and $\text{Al}_{0.3}\text{Ga}_{0.7}\text{N}$ shell are taken to be representative of their respective bulk system in the absence of defects, which has been verified experimentally \cite{mM08}. Specifically, the bandgap, electron affinity, isotropic effective mass, and dielectric constant used in this work for $\text{Al}_{x}\text{Ga}_{1-x}\text{N}$ are given by $E_{\text{g}}(x)=[3.42+2.86x-x(1-x)] \, \text{eV}$, $\chi=[5.88-0.7E_{\text{g}}(x)] \, \text{eV}$, $m^{\ast}(x)=(0.20-0.12x)m_0$, and $\varepsilon(x)=9.28-0.61x$, respectively \cite{vF03}. The specific values for GaN $(x=0)$ and $\text{Al}_{0.3}\text{Ga}_{0.7}\text{N}$ $(x=0.3)$ give rise to a Type I straddling gap heterojunction with conduction band discontinuity $\Delta E_c = 0.5 \, \text{eV}$, as shown in Fig.~\ref{F:bandDiagram}.

In GaN/AlGaN heterostructures, a spatially-dependent polarization, $\mathbf{P(r)}$, arises from two sources: (1) the spontaneous polarization, $\mathbf{P}_{\text{s}}$, due to the difference in electronegativities between GaN/AlGaN that leads to the formation of molecular dipole fields \cite{vP20}, and (2) the piezoelectric polarization, $\mathbf{P}_{\text{p}}$, due to the lattice mismatch at the epitaxially grown GaN/AlGaN interface that induces strain during thermal expansion. In both cases, a non-zero charge density emerges at the GaN/AlGaN interface due to the discontinuity in $\mathbf{P}=\mathbf{P}_{\text{s}}+\mathbf{P}_{\text{p}}$. For GaN/AlGaN crystalline systems, the spontaneous polarization can be written $\mathbf{P_\text{s}} = P_\text{s} \hat{\mathbf{z}}$, where $\hat{\mathbf{z}}$ denotes a unit vector in the $[0001]$-direction. From classical electrostatics, it follows that the interfacial charge due to spontaneous polarization at the GaN/$\text{Al}_{x}\text{Ga}_{1-x}\text{N}$ interface is given by $\sigma_\text{s} = -\nabla \cdot \mathbf{P}=(P_\text{s}^{\text{GaN}} - P_\text{s}^{\text{Al}_{x}\text{Ga}_{1-x}\text{N}}) \cos \phi$, where $\phi$ is the interfacial angle with respect to the $[0001]$-direction (a typographical error in the expression for $\sigma_\text{s}$ occurs in Ref.~\cite{bW11}). Due to this angular dependence with respect to the crystallographic axes, the interfaces in the hexagonal cross-section are all nonpolar, whereas the triangular cross-section has one polar and two semi-polar faces, as shown in Fig.~\ref{F:triPolars}. Using Vegard's law, the spontaneous polarization for $\text{Al}_{x}\text{Ga}_{1-x}\text{N}$ satisfies $P_\text{s}^{\text{Al}_{x}\text{Ga}_{1-x}\text{N}} = (1-x)P_\text{s}^{\text{GaN}} + xP_\text{s}^{\text{AlN}}$, where $P_\text{s}^{\text{GaN}} = -0.029 \, \text{C}/\text{m}^2$ and $P_\text{s}^{\text{AlN}} = -0.081 \, \text{C}/\text{m}^2$ \cite{oA00,aR06}. The specific values used for $\text{Al}_{0.3}\text{Ga}_{0.7}\text{N}$ $(x=0.3)$ give $|\sigma_\text{s}| = 0.0156 \, \text{C}/\text{m}^2$ for the polar interface (i.e., $\phi=0$), where a positive/negative interfacial charge occurs in the Ga-/N-face orientation, respectively. Conversely, the charge density on the two semi-polar interfaces at $\phi = 2 \pi / 3$ and $\phi = 4 \pi / 3$ yields $|\sigma_{\text{s}} / 2| = 0.0078 \, \text{C}/\text{m}^2$, where {the} positive/negative interfacial charge occurs on the N-/Ga-face orientation, respectively.

The change in the piezoelectric polarization $\mathbf{P}_\text{p}$ at the core--shell interface is given by $-\Delta P_{\text{p}} = \sigma_\text{p} = \varepsilon_{x^{\prime} x^{\prime}} e_{31} \cos \theta + \varepsilon_{y^{\prime} y^{\prime}} \{e_{31} \cos^3 \theta + [(e_{33}-e_{15})/2]\sin\theta \sin2\theta\} + \varepsilon_{z^{\prime} z^{\prime}}\{[(e_{31}+e_{15})/2]\sin\theta \sin2\theta + e_{33}\cos^3 \theta\} + \varepsilon_{y^{\prime} z^{\prime}}[(e_{31}-e_{33})\cos\theta \sin2\theta + e_{15}\sin\theta \cos2\theta]$, \cite{mm10} where the primed variables denote transformed coordinates in $(x',y',z')$ space, $\theta=\phi+\pi$, and the piezoelectric tensor components $e_{ij}$ are taken from Ref.~\cite{aR06}. Strain forces in the core and shell resulting from lattice mismatch have been calculated in three dimensions using several methods. For instance, the atomistic valence force-field model \cite{pK66} has been applied to hexagonal core--shell GaN/AlN NW systems \cite{kH10} and continuum elasticity theory to cylindrical core--shell NW Si/Ge geometries \cite{tT08}. Each of these prior studies indicates the interfacial strain discontinuity is similar to that of a thin film, and strain gradients within the shell are less significant than those near the interface. As such, we neglect volumetric contributions to the piezoelectric polarization of the shell and focus exclusively on the interfacial component derived from planar film expressions \cite{vF03} for each orientation considered in this work. Prior studies have also indicated that the relatively low strain in these structures allows us to safely disregard its influence on effective masses and bandgaps \cite{dC09}.

To calculate electronic properties at these nanoscale (\textit{but larger than atomistic}) length scales, we commence with the Schr\"odinger equation in the effective mass approximation:
\begin{equation}\label{E:SchrodEQ}
	\left[ -\frac{\hbar^{2}}{2} \nabla \cdot \frac{1}{m^{\ast}(\mathbf{r})} \nabla + V_T(\mathbf{r}) \right] \Psi_n(\mathbf{r}) = E_n 		\Psi_n(\mathbf{r})\text{,}
\end{equation}
where $\hbar$ is Planck's constant, $m^{\ast}(\mathbf{r})$ the spatially-dependent electron effective mass, $\Psi_n(\mathbf{r})$ the envelope wavefunction for state $n$, and $E_n$ its energy. As described in Chapter 6 of Ref.~\cite{lRM02}, the envelope wavefunction represents the slowly varying part of the total wavefunction in the presence of the periodic arrangement of atoms. The function $V_T(\mathbf{r})= V_{\text{CB}}(\mathbf{r}) + V(\mathbf{r}) + V_{\text{xc}}(\mathbf{r})$ is the sum of the conduction band edge profile $V_{\text{CB}}(\mathbf{r})$, the electrostatic potential energy $V(\mathbf{r})$, and the electron--electron exchange--correlation potential $V_{\text{xc}}(\mathbf{r})$, which we choose to be the local density approximation {(LDA) \cite{gunnar}}. For single core--shell NW geometries, $V_{\text{CB}}(\mathbf{r})$ takes the following form for hexagonal cross-sections obeying a charge-neutrality constraint:
\begin{equation}\label{E:hexInitPotential}
	V_{\text{CB}}^{\text{hex}}(x,y)=
	\begin{cases}
		0,	&\parbox[t]{.6\textwidth}{$y \leq \tfrac{\sqrt{3}}{2}c$ \& $y \leq -\sqrt{3}(x-c)$ \& $y \geq \sqrt{3}(x-c)$  \& $y \geq -\tfrac{\sqrt{3}}{2}c$ \& $y \geq -\sqrt{3}(x+c)$ \& $y \leq \sqrt{3}(x+c)$;}\\
		\Delta E_c,	&\text{otherwise.}
	\end{cases}
\end{equation}
For hexagonal cross-sections constrained with an externally-pinned Fermi level, $V_{\text{CB}}(\mathbf{r})$ takes the following form for hexagonal cross-sections:
\begin{equation}\label{E:hexInitPotential_fermi}
	V_{\text{CB}}^{\text{hex}}(x,y)=
	\begin{cases}
		-\Delta E_c,	&\parbox[t]{.6\textwidth}{$y \leq \tfrac{\sqrt{3}}{2}c$ \& $y \leq -\sqrt{3}(x-c)$ \& $y \geq \sqrt{3}(x-c)$  \& $y \geq -\tfrac{\sqrt{3}}{2}c$ \& $y \geq -\sqrt{3}(x+c)$ \& $y \leq \sqrt{3}(x+c)$;}\\
		0,	&\text{otherwise,}
	\end{cases}
\end{equation}
and the following form for triangular cross-sections:
\begin{equation}\label{E:triInitPotential}
	V_{\text{CB}}^{\text{tri}}(x,y)=
	\begin{cases}
		-\Delta E_c,	&\text{$y \geq -\tfrac{\sqrt{3}}{6}c$ \& $y \leq \sqrt{3}x+\tfrac{\sqrt{3}}{3}c$ \& $y \leq -\sqrt{3}x+\tfrac{\sqrt{3}}{3}c$;}\\
		0,	&\text{otherwise,}
	\end{cases}
\end{equation}
where we have chosen a coordinate system such that the $z$-axis passes through the geometric center of NW cross-section area, the core side length, $c$, is depicted in Figs.~\ref{F:hexCross} and ~\ref{F:triCross}, and $\Delta E_c=0.50$ eV is the ${\text{Al}}_{0.3}{\text{Ga}}_{0.7}\text{N}$ conduction band edge discontinuity depicted in Fig.~\ref{F:bandDiagram}. {For core-multishell NW geometries examined in this work, each separate region is described with a $V_{\text{CB}}(\mathbf{r})$ expression having the same functional form as Eqs.~\eqref{E:hexInitPotential} or ~\eqref{E:hexInitPotential_fermi}.} As discussed further in Section 3, it is important to note that the HADOKEN code sets the zero of energy at the minimum of the conduction band edge for NWs obeying a charge-neutrality constraint. In contrast, the zero of energy is set at the outer shell edge for NWs constrained with an externally-pinned Fermi level.

For computational convenience, the Schr\"odinger equation in the HADOKEN code is converted to a dimensionless form using the following reduced variables: $\tilde{x}=x/\ell_0$, $\tilde{y}=y/\ell_0$, $\tilde{z}=z/\ell_0$, and $E=\epsilon \mathcal{C}$, where $\ell_0$ is a characteristic length scale, $\mathcal{C} = \hbar^2/2 m_0 \ell_{0}^{2}$, and $m_0$ is the electron rest mass. Within the HADOKEN code, $\ell_0$ is set to 10 nm, which correspondingly sets the energy scaling factor $\mathcal{C}$ to be 0.381\,meV. Assuming translational invariance along the $z$-axis, the envelope wavefunction in Eq.~\eqref{E:SchrodEQ} can be factored as $\Psi_{n}(\mathbf{r}) = {e^{i k z} \psi_{n}(x,y)}/{\sqrt {L}} \mspace{1mu}$, where $L$ is a normalization factor along the length of the NW, and $k$ is the wavevector along the NW axis. Expressing $\Psi_{n}(\mathbf{r})$ in terms of reduced variables requires some care, since the HADOKEN code numerically calculates the wavefunctions $\psi_{n}(\tilde{x},\tilde{y})$ and normalizes them over the NW cross-section \textit{in the reduced coordinates $\tilde{x}$ and $\tilde{y}$}. Quantum mechanics requires $\iint dx\,dy\: |\psi_{n}(x,y)|^2 = 1$ in the \textit{unscaled coordinates}; however, the HADOKEN code uses the normalization convention $\iint d\tilde{x}\,d\tilde{y}\: |\psi_{n}(\tilde{x},\tilde{y})|^2 = 1$ in reduced coordinates (the normalization constant  is calculated as one of the outputs in \texttt{normalize\char`_and\char`_sqrt\char`_m\char`_triangular.m}). To satisfy both of these constraints we must define $\psi_{n}(x,y)=\psi_{n}(\tilde{x},\tilde{y})/\ell_0$. Therefore, in terms of the reduced coordinates, the envelope wavefunction in Eq.~\eqref{E:SchrodEQ} becomes
\begin{equation}\label{E:red_envelope_wfn}
\Psi_{n}(\mathbf{r}) = \frac{1}{\ell_0 \! \sqrt {L}}e^{i k \ell_0 \tilde{z}} \psi_{n}(\tilde{x},\tilde{y}) \text{,}
\end{equation}
where $|\psi_{n}(\tilde{x},\tilde{y})|^2$ \textit{is dimensionless and its integral over $\tilde{x}$ and $\tilde{y}$ is normalized to unity}.

Substituting Eq.~\eqref{E:red_envelope_wfn} into  Eq.~\eqref{E:SchrodEQ} and restricting our study to only electronic properties at the Gamma point (i.e., $k=0$) gives the following reduced, two-dimensional Schr\"odinger equation:
\begin{equation}\label{E:redSchrodEQ}
	\left[ -\frac{\partial}{\partial\tilde{x}}  \frac{m_0}{m^{\ast}(\tilde{x},\tilde{y})} 					\frac{\partial}{\partial\tilde{x}} -\frac{\partial}{\partial\tilde{y}}  \frac{m_0}{m^{\ast}(\tilde{x},\tilde{y})} 					\frac{\partial}{\partial\tilde{y}}+ \frac{V_T(\tilde{x},\tilde{y})}{\mathcal{C}}  \right] \psi_{n}(\tilde{x},\tilde{y}) = \epsilon_{n} \psi_{n}(\tilde{x},\tilde{y}) \text{,}
\end{equation}
where $m^{\ast}(\tilde{x},\tilde{y})$ and $V_T(\tilde{x},\tilde{y})$ are only functions of $\tilde{x}$ and $\tilde{y}$ due to the translational invariance along the $z$-axis. In the HADOKEN code,  $m^{\ast}(\tilde{x},\tilde{y})$ is computed using the \texttt{heaviside\char`_core\char`_schrod.m} routine which is subsequently used as input as the ``c" coefficient in the MATLAB PDE Toolbox \texttt{pdeeig} command. In addition, the HADOKEN code uses Dirichlet boundary conditions for Eq.~\eqref{E:redSchrodEQ} where  $\psi_n(\tilde{x},\tilde{y})$ is set to zero at the outer shell boundary to prevent any electron leakage outside the NW.
The electrostatic potential energy, $V(\tilde{x},\tilde{y})$, satisfies Poisson's equation, which, in cgs units, is given by:
\begin{equation}\label{E:redPoissEQ2}
        \begin{split}
	\left[ \frac{\partial}{\partial\tilde{x}} \varepsilon(\tilde{x},\tilde{y}) \frac{\partial}{\partial\tilde{x}} +\frac{\partial}{\partial\tilde{y}}  \varepsilon(\tilde{x},\tilde{y}) \frac{\partial}{\partial\tilde{y}} \right] V(\tilde{x},\tilde{y}) &= 4 \pi \ell_{0}^2 |e| \left[ \rho_{D}(\tilde{x},\tilde{y}) + \rho_{e}(\tilde{x},\tilde{y}) + \nabla \cdot \mathbf{P}\right] \\
	&=S_D(\tilde{x},\tilde{y})+S_e(\tilde{x},\tilde{y})+4 \pi \ell_{0}^2 |e|\nabla \cdot \mathbf{P},
	    \end{split}
\end{equation}
where $e$ is the charge of an electron, $\varepsilon(\tilde{x},\tilde{y})$ is the spatially-dependent static dielectric constant, $\rho_{D}(\tilde{x},\tilde{y})$ is the charge density arising from the presence of ionized donors, $\rho_{e}(\tilde{x},\tilde{y})$ is the electron density, $\mathbf{P}$ $(=\mathbf{P}_\text{s}+\mathbf{P}_\text{p})$ is the total polarization source term discussed previously (which is only relevant for the Ga-face or N-face triangular nanowires), and $S_D(\tilde{x},\tilde{y})$ and $S_e(\tilde{x},\tilde{y})$ are defined as source terms due to the ionized donors and electron density, respectively. In the HADOKEN code, $\varepsilon^{\ast}(\tilde{x},\tilde{y})$ is computed using the \texttt{heaviside\char`_core\char`_poiss.m} routine which is subsequently used as input for the ``\texttt{c}" coefficient in the MATLAB PDE Toolbox \texttt{assempde} command. It is important to note that $\nabla$ in Eq.~\eqref{E:redPoissEQ2} is the two-dimensional gradient operator $(=[\partial/\partial x,\,\partial/\partial y])$ in regular (\textit{not reduced}) variables. The $S_D(\tilde{x},\tilde{y})$ and $S_e(\tilde{x},\tilde{y})$ source terms in Eq.~\eqref{E:redPoissEQ2} are discussed separately below, and the $\nabla \cdot \mathbf{P}$ term is discussed in greater detail in Section 4.2. {Finally, the HADOKEN code can utilize either Dirichlet or Neumann boundary conditions for Eq.~\eqref{E:redPoissEQ2} where either $V(\tilde{x},\tilde{y})$ or its derivative, respectively, are set to zero at the outer shell boundary.}

For computational convenience, both $S_{D}(\tilde{x},\tilde{y})$ and $S_{e}(\tilde{x},\tilde{y})$ are expressed in terms of electron and donor number densities, $n_e$ and $n_D$, within the HADOKEN code as 
\begin{equation}\label{E:chrgDensities_D}
        S_{D}(\tilde{x},\tilde{y}) = 4 \pi \ell_{0}^2 |e|^2  n_D(\tilde{x},\tilde{y}) \Theta \left[ V_T(\tilde{x},\tilde{y}) - E_F \right]\text{,}
\end{equation}
and
\begin{equation}\label{E:chrgDensities_e}
            S_{e}(\tilde{x},\tilde{y}) = -4 \pi \ell_{0}^2 |e|^2  n_e(\tilde{x},\tilde{y})\text{.}
\end{equation}
The Heaviside step-function, $\Theta$, determines the depletion region, which is the cross-sectional area of the NW where donor ionization can take place. Discussed further at the end of this section, the depletion region denotes areas of the NW where the donor electrons have energies larger than the Fermi energy, $E_F$, for ionization to occur (cf.  Chapter 6 of Ref. \cite{lRM02}). Within HADOKEN, the number density, $n_D$, is scaled by a typical carrier density of $n_{D,18}=10^{18}\,{\text{cm}}^{-3}$, allowing us to concisely express the source term, $S_D$, in units of eV in the \texttt{n\char`_D\char`_func.m} m-file to give:
\begin{equation}\label{source_donor_eV}
	S_D(\tilde{x},\tilde{y}) = (1.80951 \, \text{eV}) \left[ \frac{n_D(\tilde{x},\tilde{y})}{n_{D,18}}\right] \Theta \left[ V_T(\tilde{x},\tilde{y}) - E_F \right]  \text{.}	
\end{equation}

Similarly, the source term due the electron density can be written as
\begin{equation}\label{electron_donor_eV}
	S_e(\tilde{x},\tilde{y}) = -(1.80951 \, \text{eV}) \ell_{0}^3 n_e(\tilde{x},\tilde{y}) \text{.}
\end{equation}
The electron number density, $n_e$, at temperature $T$ is obtained by summing over the total number of occupied states:
\begin{equation}\label{E:elecNumDensity1}
	n_e(\tilde{x},\tilde{y}) = 2 \sum_{n, \mspace{1mu} n_z} |\Psi_n(x,y,z)|^2 \mspace{1mu} f(E,E_F,T) \text{,}	
\end{equation}
where the factor of 2 on the right side accounts for the spin degeneracy of each energy level. The sum in Eq.~\eqref{E:elecNumDensity1} extends over the quantum numbers $n$ and $n_z$, which correspond to quantization across the NW cross-section and axis, respectively. The Fermi distribution, $f(E,E_F,T)$ in Eq.~\eqref{E:elecNumDensity1}, is given by
\begin{equation}\label{E:fermiDirac}
	f(E,E_F,T) =
	\begin{cases}
		\frac{1}{e^{(E - E_F) / k_{B}T} + 1},	&\text{for $T \neq 0\,\text{K}$;}\\
		\Theta(E_F - E),			        	&\text{for $T = 0\,\text{K}$;}
	\end{cases}
\end{equation}
where $k_B$ is Boltzmann's constant. As described further below, we only consider the $T=0\,\text{K}$ case, since the resulting integrals over $\Theta(E_F - E)$ have a closed-form, analytic solution that can be efficiently computed over numerous self-consistent iterations within the HADOKEN code.

The electrons have a continuous energy spectrum for motion along the NW axis, allowing the sum over $n_z$ to be rewritten as a continuous integral. Accordingly, Eq.~\eqref{E:red_envelope_wfn} can be substituted into Eq.~\eqref{E:elecNumDensity1} to give
\begin{equation}\label{E:elecNumDensity2}
	n_e(\tilde{x},\tilde{y}) = \frac{2}{\ell_0^2 L} \sum_{n} |\psi_n(\tilde{x},\tilde{y})|^2 \int dn_z \: f(E,E_F,T)\text{.}	
\end{equation}
Using the relation $k_z = 2 \pi n_z / L$, Eq.~\eqref{E:elecNumDensity2} may be converted to a momentum-space integral over $k_z$:
\begin{equation}\label{E:elecNumDensity3}
	n_e(\tilde{x},\tilde{y}) = \frac{1}{\pi \ell_0^2} \sum_{n} |\psi_n(\tilde{x},\tilde{y})|^2 \int dk_z \: f(E,E_F,T)\text{.}	
\end{equation}
Considering only the $T=0\,\text{K}$ case in the Fermi distribution with the relations $E=E_z+E_n$, $E_z=\hbar^2k_z^2/2 m^{\ast}(\tilde{x},\tilde{y})$, and $dk_z = \sqrt{{m^{\ast}(\tilde{x},\tilde{y})}/{2 \hbar^2 E_z}} \: dE_z$ (the second expression assumes that the effective mass can be closely approximated as a scalar and factored through the spatial derivatives appearing in Schr\"odinger's equation) gives
\begin{equation}\label{E:elecNumDensity4}
	n_e(\tilde{x},\tilde{y}) = \frac{1}{\pi \ell_0^2} \sum_n |\psi_n(\tilde{x},\tilde{y})|^2 \int_{0}^{\infty} dE_z \, \sqrt{\tfrac{m^{\ast}(\tilde{x},\tilde{y})}{2\hbar^2 E_z}} \,
	\Theta(E_f - E_z - E_n) \text{.}
\end{equation}
Due to the Heaviside function, $\Theta(E_f-E_z-E_n)$, the integral in Eq.~\eqref{E:elecNumDensity4} is only nonzero when when $E_z < E_F - E_n$, which gives
\begin{equation}\label{E:elecNumDensity5}
 \begin{split}
	n_e(\tilde{x},\tilde{y}) &= \frac{1}{\pi \ell_0^2} \sum_n |\psi_n(\tilde{x},\tilde{y})|^2 \int_{0}^{E_F - E_n} dE_z \, \sqrt{\tfrac{m^{\ast}(\tilde{x},\tilde{y})}{2\hbar^2 E_z}} \\
	&=\frac{1}{\pi \ell_0^2} \sum_n |\psi_n(\tilde{x},\tilde{y})|^2 \sqrt{\tfrac{2m^{\ast}(\tilde{x},\tilde{y})(E_F - E_n)}{\hbar^2}}\\
	&=\frac{1}{\pi \ell_0^3} \sum_n |\psi_n(\tilde{x},\tilde{y})|^2 \sqrt{\tfrac{m^{\ast}(\tilde{x},\tilde{y})(\epsilon_F-\epsilon_n)}{m_0}}\text{,}
	 \end{split}
\end{equation}
where we have used the reduced variable $E=\epsilon \mathcal{C}$ in the last step. Substituting Eq.~\eqref{E:elecNumDensity5} into Eq.~\eqref{electron_donor_eV} allows us to concisely express the source term, $S_e$, in units of eV in the \texttt{n\char`_e\char`_func.m} m-file to give:
\begin{equation}\label{electron_donor_eV_simple}
	S_e(\tilde{x},\tilde{y}) = -\frac{1.80951\,\text{eV}}{\pi} \sum_n |\psi_n(\tilde{x},\tilde{y})|^2 \sqrt{\tfrac{m^{\ast}(\tilde{x},\tilde{y})(\epsilon_F-\epsilon_n)}{m_0}}  \text{.}
\end{equation}
Within the HADOKEN code, the ${\pi}^{-1} \! \sqrt{m^{\ast}(\tilde{x},\tilde{y})/{m_0}}$ ``prefactor" term is calculated in \texttt{n\char`_e\char`_prefactor.m}, the summation $\sum_n |\psi_n(\tilde{x},\tilde{y})|^2 \sqrt{\epsilon_F-\epsilon_n}$ is computed in \texttt{psi\char`_sqrt\char`_eps\char`_summation.m}, and \texttt{n\char`_e\char`_func.m} utilizes the output of the previous two m-files to ultimately calculate $S_e$ as input to the Poisson equation.

\begin{figure}[H]
	\centering
	\begin{subfigure}[t]{0.48\textwidth}
		\centering
		\includegraphics[width=\textwidth]{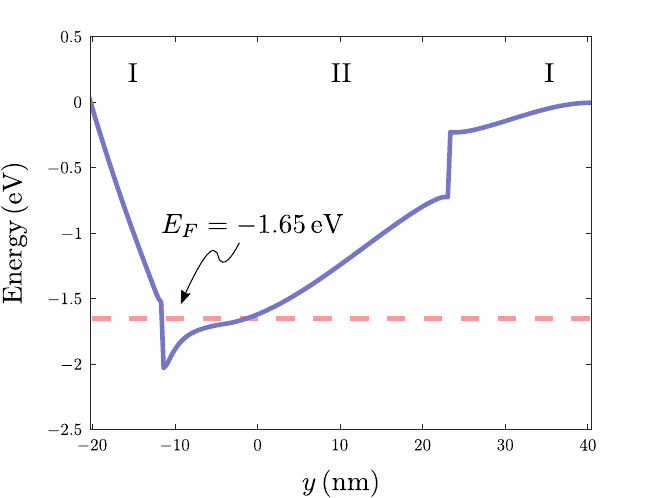}
		\caption{}
		\label{F:facePinBandBend1}
	\end{subfigure}
	\hfill
	\begin{subfigure}[t]{0.48\textwidth}
		\centering
		\includegraphics[width=\textwidth]{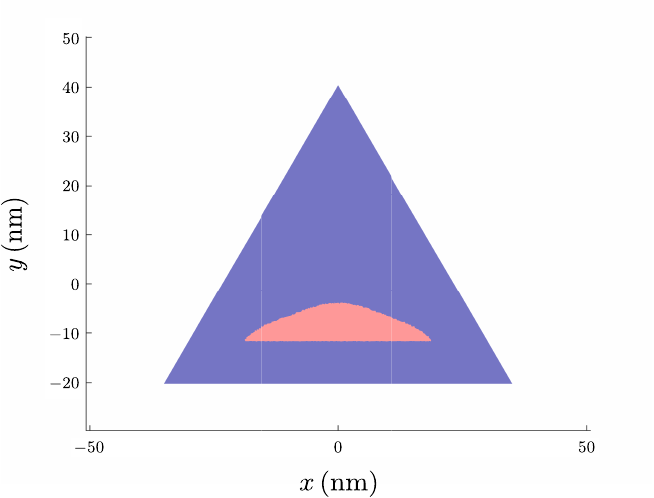}
		\caption{}
		\label{F:facePinDepletionRegion}
	\end{subfigure}
	\caption{(a)\,Band-bending and Fermi level for a (0001) Ga-face triangular core--shell nanowire, and (b)\,corresponding cross-section of the NW showing the depletion region colored in purple.}
	\label{F:facePinBBandDR}
\end{figure}

Thus far, neither the Fermi level nor the NW depletion region has been specified in our computational description. To more concretely describe these concepts, we first describe the Fermi level and depletion region for the (0001) Ga-face triangular core--shell nanowire depicted in Fig.~\ref{F:facePinBBandDR}. Due to surface states in these systems (discussed further in Section~\ref{ss:triResults}), the Fermi level is pinned $1.65\,\text{eV}$ below the conduction band edge, as shown in Fig.~\ref{F:facePinBandBend1} \cite{gK05}. The corresponding NW depletion region, colored purple in Fig.~\ref{F:facePinDepletionRegion}, contains the positively-charged (ionized) donors that provide carriers in the NW (i.e., the purple-colored regions contribute a positive-valued dopant density, $n_D$).

For NWs without a Fermi-pinning constraint, HADOKEN can use a charge neutrality condition (discussed further in Section~\ref{methods}) in combination with the solution of the depletion region to calculate $\epsilon_F$. The charge neutrality condition requires the total number of positive and negative charges over the entire NW to balance:
\begin{equation}\label{E:chrgNeutral}
	\iint d\tilde{x} \, d\tilde{y} \: n_D(\tilde{x},\tilde{y}) = \iint d\tilde{x} \, d\tilde{y} \: n_e(\tilde{x},\tilde{y}) \text{,}
\end{equation}
where $n_D$ and $n_e$ are described in Eqs.~\eqref{E:chrgDensities_D} and \eqref{E:elecNumDensity5}. Since $n_e$ is a function of $\epsilon_F$ (see Eq.~\eqref{E:elecNumDensity5}) and $n_D$ is a function of the depletion region, the Fermi energy is calculated by solving Eq.~\eqref{E:chrgNeutral} using a standard root-finding procedure in the HADOKEN m-files \texttt{charge\char`_neutral.m} and \texttt{find\char`_epsilon\char`_F.m} The \texttt{normalize\char`_and\char`_sqrt\char`_m\char`_triangular.m} routine calculates $ \iint d\tilde{x} \, d\tilde{y} \, |\psi_n(\tilde{x},\tilde{y})|^2 \sqrt{{m^{\ast}(\tilde{x},\tilde{y})}/{m_0}}$ for each of the $n$ wavefunctions, which is needed as input to \texttt{find\char`_epsilon\char`_F.m}. The depletion region itself is calculated/stored in the variable \texttt{heaviside\char`_n\char`_D} within the HADOKEN code. Specifically, the geometric regions of the NW that satisfy $V_{T}(\tilde{x},\tilde{y}) \geq E_F$ become ionized and contribute a positive-valued dopant density, $n_D$, that is taken into account in the charge neutrality condition of Eq.~\eqref{E:chrgNeutral}. The procedure for satisfying charge neutrality is carried out during each iterative cycle until self-consistency in the total potential, $V_{T}(\tilde{x},\tilde{y})$, is reached (described further in the following section).

\section{Additional Numerical and Implementation Details}\label{methods}
\begin{figure}[H]
	\centering
	\includegraphics[width=0.75\textwidth]{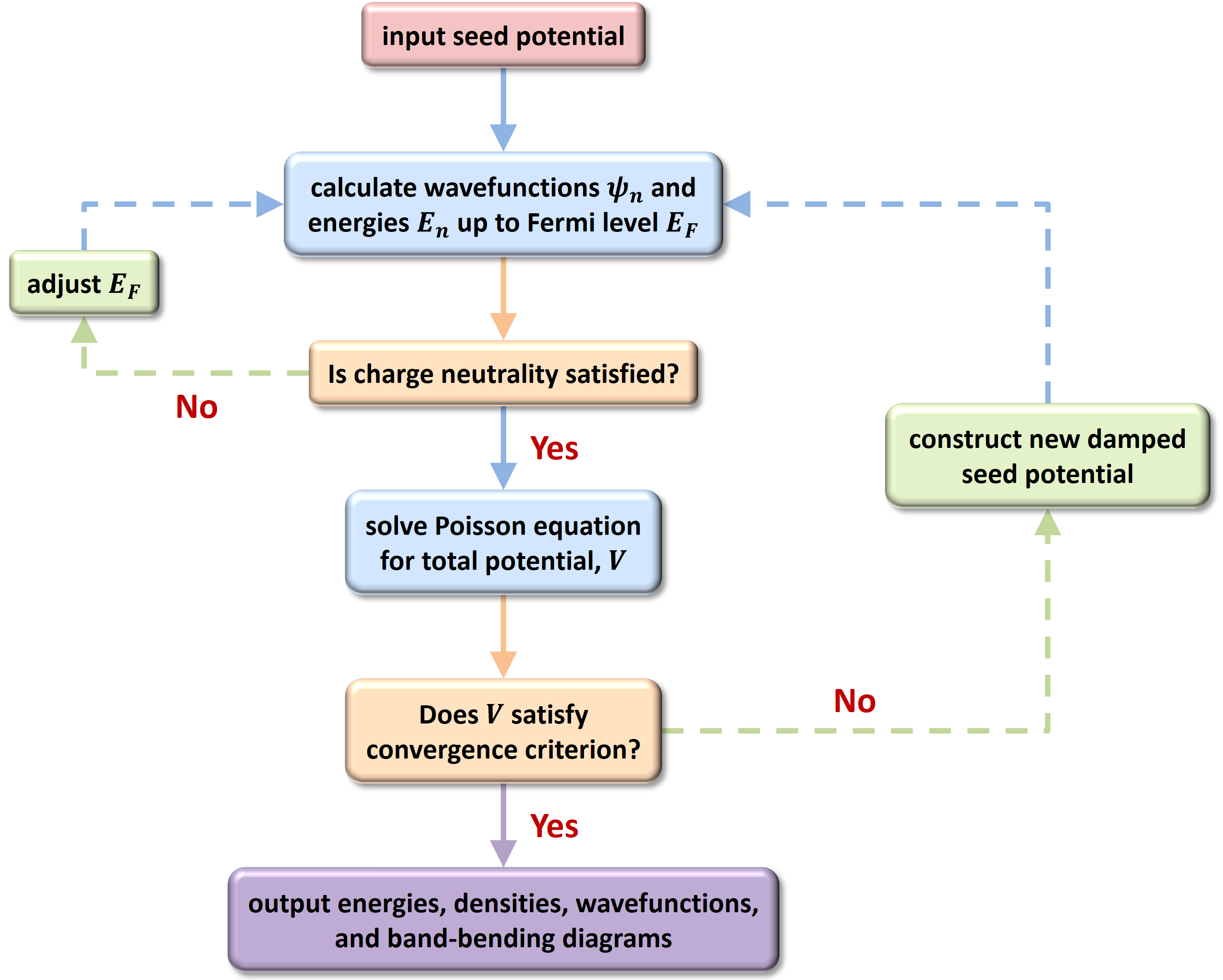}
	\caption{Algorithmic flowchart of the HADOKEN code for nanowires obeying a charge-neutrality constraint. For nanowires constrained with an externally-pinned Fermi level, the charge-neutrality decision block is bypassed, and the Poisson equation is solved immediately after the wavefunctions are calculated.}
	\label{F:flowChart}
\end{figure}

{The HADOKEN source code is distributed as a collection of MATLAB m-files in the following four separate, self-descriptive folders:
\newline
\newline
\texttt{hexagonal\char`_charge\char`_neutral\char`_coreshell}
\newline
\texttt{hexagonal\char`_fermi\char`_pinning\char`_coreshell}
\newline
\texttt{triangular\char`_Ga\char`_face\char`_coreshell}
\newline
\texttt{triangular\char`_N\char`_face\char`_coreshell}
\newline}
\newline
The most computationally intensive portions in these m-files utilize the MATLAB Partial Differential Equation Toolbox~\cite{PDE_toolbox} to calculate self-consistent electronic wavefunctions, energies, densities, and band-bending diagrams for hexagonal and triangular core--shell nanowires. The flowchart depicted in Fig.~\ref{F:flowChart} summarizes the overall algorithmic processes within HADOKEN, which are described in extensive detail in the following paragraphs. {The \texttt{set\char`_input\char`_parameters.m} and \texttt{set\char`_doping\char`_density.m} routines allow the user to input parameters specifying the desired core/shell side lengths, conduction band edge energies, effective masses, dielectric constants, mesh resolution, and doping density function, respectively. These quantities are then used by the \texttt{main\char`_scf\char`_dirichlet.m} and \texttt{main\char`_scf\char`_neumann.m} routines, which initiate the HADOKEN code with either Dirichlet or Neumann boundary conditions for $V(\tilde{x},\tilde{y})$. As sample input, the material parameters for GaN and $\text{Al}_{0.3}\text{Ga}_{0.7}\text{N}$ have been provided in \texttt{set\char`_input\char`_parameters.m} in the global variables \texttt{vector\char`_of\char`_V0}, \texttt{vector\char`_of\char`_masses}, and \texttt{vector\char`_of\char`_eps}, which represent the conduction band discontinuities, effective masses, and dielectric constants of each of the nanowire regions, respectively. Researchers interested in other material compositions can simply replace these numerical values in the \texttt{set\char`_input\char`_parameters.m} file to enable self-consistent simulations for other materials.}

\subsection{Finite Element Mesh Generation}\label{ss:FEMgeneration}
\begin{figure}[H]
	\centering
	\begin{subfigure}[t]{0.48\textwidth}
		\centering
		\includegraphics[width=\textwidth]{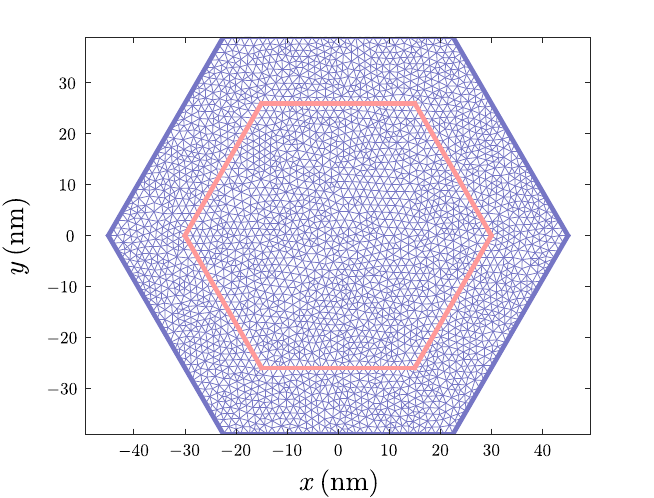}
		\caption{}
		\label{F:hexMesh}
	\end{subfigure}
	\hfill
	\begin{subfigure}[t]{0.48\textwidth}
		\centering
		\includegraphics[width=\textwidth]{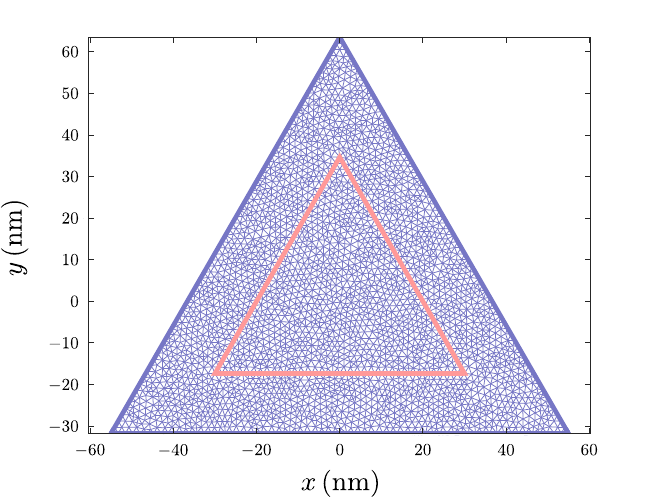}
		\caption{}
		\label{F:cornerPinMesh}
	\end{subfigure}
	\caption{Representative finite element meshes automatically generated by HADOKEN for (a) hexagonal and (b) triangular core--shell NWs. The pink lines delineate the core region of each geometry.}
	\label{F:meshes}
\end{figure}

{With the global variable \texttt{vector\char`_of\char`_side\char`_lengths} properly defined in the \texttt{set\char`_input\char`_parameters.m} routine,} the \texttt{mesh\char`_coreshell.m} m-file generates a Delaunay-triangulated grid of points that discretizes the NW cross-sectional geometry using the built-in \texttt{initmesh} MATLAB function \cite{webRef}. Specifically, the \texttt{initmesh} function utilizes a decomposed geometry matrix, \texttt{g}  \cite{webRef_geometry}, and outputs the matrices \texttt{p}, \texttt{e}, and \texttt{t} for point, edge, and triangular mesh data (\texttt{p} and \texttt{e} are stored as global variables that are used by several other routines in HADOKEN). For simplicity, the same mesh grid is used for both the Schr\"odinger and Poisson equations. The average side length of the individual triangles forming this finite element mesh is computed and stored in the global variables \texttt{avg\char`_side\char`_length\char`_schrod} and \texttt{avg\char`_side\char`_length\char`_poiss}, which are used by several of the other MATLAB m-files. Fig.~\ref{F:meshes} shows representative hexagonal/triangular cross-sectional geometries and finite element grids that are automatically plotted by the the built-in \texttt{pdemesh} MATLAB function in HADOKEN. {It is also important to note that researchers interested in different NW cross-sections can construct a customized geometry matrix, \texttt{g} (see Ref.~\cite{webRef_geometry} for further details), which can use the same iteration scheme (discussed further below) in the HADOKEN code for their own self-consistent calculations.}

\subsection{Initial Guess for the Potential}\label{ss:initGuessPotential}
With the material parameters and mesh data properly defined/computed, HADOKEN initializes the self-consistent procedure by providing a zeroth-order guess for the total potential, $V_T(\tilde{x},\tilde{y})$. For hexagonal geometries not containing polarization source terms, this guess potential is provided by the \texttt{V\char`_conduction\char`_band.m} m-file, which contains only the bare conduction band edge profile, $V_{\text{CB}}^{\text{hex}}(\tilde{x},\tilde{y})$, given by Eqs.~\eqref{E:hexInitPotential} or \eqref{E:hexInitPotential_fermi}. As mentioned previously in Section~\ref{theory}, \texttt{V\char`_conduction\char`_band.m} sets the zero of the potential energy at the minimum of the conduction band edge for NWs obeying a charge-neutrality constraint. For NWs constrained with an externally-pinned Fermi level, the zero of energy is set at the outer shell edge. Since $V_{\text{CB}}^{\text{hex}}(\tilde{x},\tilde{y})$ is defined piecewise in the core/shell regions, the general-purpose \texttt{heaviside\char`_core\char`_schrod.m} m-file---which returns a value of 1 for inputted $(\tilde{x},\tilde{y})$ pairs that lie within the core region of the nanowire and 0 otherwise---is used to construct an appropriately scaled conduction band edge profile in \texttt{V\char`_conduction\char`_band.m}.

For both of the Ga- and N-face triangular geometries that contain spontaneous and piezoelectric polarization source terms, the guess potential is obtained by numerically solving the dimensionless Poisson equation in Eq.~\eqref{E:redPoissEQ2} without the $S_e(\tilde{x},\tilde{y})$ source term. In addition, the $S_D(\tilde{x},\tilde{y})$ source term in Eq.~\eqref{E:redPoissEQ2} is approximated as a constant for this first iteration of the Poisson equation to obtain the guess potential. As discussed in Section~\ref{theory}, the spontaneous polarization source term yields a charge density at each of the three GaN/AlGaN interfaces for the triangular cross-section. From classical electrostatics, this interfacial charge density is formally represented by a Dirac delta function at each of the three interfaces. For example, the charge density at the polar interface for the Ga-face orientation is given by $\sigma_\text{s}=(0.0156 \, \text{C}/\text{m}^2) \, \delta(\tilde{y}+c\sqrt{3}/6)$ within the domain $-c/2 \leq \tilde{x} \leq c/2$, where $c$ is the core side length. This delta function charge distribution is implemented in the \texttt{rho\char`_semipolar\char`_inside.m} m-file as a Gaussian function of the form $\rho(\tilde{x},\tilde{y}) = \sigma_{\text{s}} \exp{[-(\tilde{y}+c\sqrt{3}/6)^2 / a^2]} / {a \! \sqrt{\pi}}$ within the domain $-c/2 \leq \tilde{x} \leq c/2$, where $a$ is twice the average side length of the triangular finite element mesh (i.e., $a = 2 \cdot \texttt{avg\char`_side\char`_length\char`_poiss}$). This charge density is then stored in the variable \texttt{rho\char`_bottom\char`_inside}. While the numerical scheme for incorporating the charge density at the polar interface is relatively straightforward, implementing the surface charge for the semi-polar interfaces requires additional care since the Delaunay triangulation procedure does not create a symmetric grid. To this end, a rotation matrix is used in \texttt{rho\char`_semipolar\char`_inside.m} to first rotate the $(\tilde{x},\tilde{y})$ coordinates of the finite element grid by 120$^{\circ}$ counterclockwise so that one semi-polar interface now lies along the horizontal line $\tilde{y}=-c\sqrt{3}/6$. A Gaussian having the same functional form as the one described previously (except with $\sigma_\text{s}$ replaced with $-\sigma_\text{s}/2$) is then applied at the $\tilde{y}=-c\sqrt{3}/6$ line within the domain $-c/2 \leq \tilde{x} \leq c/2$. The charge density is stored in the appropriate $(\tilde{x},\tilde{y})$ locations within the variable \texttt{rho\char`_left\char`_inside}. The same procedure is repeated for the other semi-polar interface (except with a 120$^{\circ}$ clockwise rotation) and the charge density is stored in the variable \texttt{rho\char`_right\char`_inside}. The three charge densities are finally added together and returned as an output variable by \texttt{rho\char`_semipolar\char`_inside.m}.

The piezoelectric polarization source term is implemented in a manner similar to the spontaneous polarization---a Gaussian function is used to approximate the interfacial charge density and rotation matrices are used to place the charge density at each of the three GaN/AlGaN interfaces. The only difference is that the charge density due to the piezoelectric polarization is given by the analytic expression for $\sigma_{\text{p}}$ discussed in Section~\ref{theory}. The analytic expression and charge density at each of the three GaN/AlGaN interfaces is computed by the \texttt{piezo\char`_analytic.m} and \texttt{rho\char`_strain\char`_analytic.m} m-files, respectively.

With the source terms properly computed, the dimensionless Poisson equation can now be solved for the guess potential for the Ga- and N-face triangular geometries. Specifically, the $S_D(\tilde{x},\tilde{y})$ and  $\nabla \cdot \mathbf{P}$ source terms are used as input for the ``\texttt{f}" coefficient in the MATLAB PDE Toolbox \texttt{assempde} function \cite{assempde}. Since the dielectric constant has a spatial dependence in the Poisson equation, the \texttt{heaviside\char`_core\char`_poiss.m} m-file (which returns a value of 1 for inputted $(\tilde{x},\tilde{y})$ values that lie within the core region of the nanowire and 0 otherwise) is used to construct an appropriately scaled $\varepsilon^{\ast}(\tilde{x},\tilde{y})$ term, which is used as the ``\texttt{c}" coefficient in the \texttt{assempde} solver. {Finally, a boundary condition matrix, \texttt{b}, which enforces Dirichlet or Neumann boundary conditions for $V(\tilde{x},\tilde{y})$ at the outer shell boundary, is used as input to \texttt{assempde}. Researchers interested in applying different boundary conditions can simply modify the boundary condition matrix, \texttt{b}, to suit their specific needs (see Ref. \cite{assempde_boundary} for the MATLAB documentation on modifying this variable).} The electrostatic potential energy, $V(\tilde{x},\tilde{y})$, is then computed and stored in the global variable \texttt{V\char`_poisson}.

\subsection{Initial Schr\"odinger Equation}\label{ss:initSchroEq}
With the initial guess for the potential calculated, HADOKEN computes initial wavefunctions and energies from the Schr\"odinger equation (Eq.~\eqref{E:redSchrodEQ}) using the Arnoldi algorithm \cite{wA51} within the MATLAB PDE Toolbox \texttt{pdeeig} function \cite{pdeeig}. For both charge neutral and Fermi-pinned hexagonal geometries, the output of the previously discussed \texttt{V\char`_conduction\char`_band.m} is used as input for the ``\texttt{a}" coefficient in \texttt{pdeeig}. For both the Ga- and N-face triangular geometries, the ``\texttt{a}" coefficient uses the output of the \texttt{V\char`_total\char`_piezo.m} m-file, which calculates the sum of \texttt{V\char`_poisson} (containing the spontaneous and piezoelectric polarization contributions) and the conduction band edge profile, $V_{\text{CB}}^{\text{tri}}(x,y)$, given by Eq.~\eqref{E:triInitPotential}. Since the effective mass has a spatial dependence in the Schr\"odinger equation, the \texttt{heaviside\char`_core\char`_schrod.m} m-file (which has a similar functionality as the \texttt{heaviside\char`_core\char`_poiss.m} m-file discussed above) is used to construct an appropriately scaled ${m_0}/{m^{\ast}(\tilde{x},\tilde{y})}$ term, which is used as the ``\texttt{c}" coefficient in the \texttt{pdeeig} PDE solver. A two-element vector, \texttt{r}, containing the range of eigenvalues to compute is also used by the \texttt{pdeeig} command: for hexagonal cross-sections obeying a charge-neutrality constraint, \texttt{r} is initially set to $[0,\mathcal{C}]$; however, for NWs with a Fermi-pinning constraint, the lower limit of \texttt{r} is initially set to the minimum of the electrostatic potential energy, and the upper limit is set to $\mathcal{C}$ or $10\mathcal{C}$ above the previous lower limit for Ferm-pinned hexagonal or triangular geometries, respectively. A boundary condition matrix, \texttt{b}, \cite{assempde_boundary} which sets the wavefunctions to zero at the outer shell edge, is used as input to \texttt{pdeeig}. Finally, at least two wavefunctions and their associated energy levels are computed by incrementally changing the lower and upper limit of the \texttt{r} vector in subsequent function calls to \texttt{pdeeig}. The \texttt{normalize\char`_and\char`_sqrt\char`_m\char`_triangular.m} m-file calculates the normalization constant and $\iint d\tilde{x} \, d\tilde{y} \, |\psi_n(\tilde{x},\tilde{y})|^2 \sqrt{{m^{\ast}(\tilde{x},\tilde{y})}/{m_0}}$ for each of the $n$ wavefunctions (the latter expression is used evaluate $n_e(\tilde{x},\tilde{y})$ in Eq.~\eqref{E:elecNumDensity5}).

\subsection{Calculation of the Fermi Level and Boundary Conditions}\label{ss:fermiLevel&BCs}
With the first two normalized wavefunctions and energies computed from the previous step, the HADOKEN code can now proceed to calculate the Fermi level, $\epsilon_F$. For both the Ga- and N-face triangular geometries, the AlGaN/vacuum interface contains a high density of surface states that counterbalance the large spontaneous polarization charge generated at the interface \cite{gK05}. This results in the Fermi level being pinned in the AlGaN bandgap near $-1.65$ eV, which sets the upper limit of the \texttt{r} vector in subsequent function calls to \texttt{pdeeig}. {For other user-defined geometries with a Fermi-pinning constraint, the Fermi level is specified by the user in the \texttt{set\char`_input\char`_parameters.m} routine.}

For NWs obeying a charge-neutrality constraint, Eq.~\eqref{E:chrgNeutral} is used to calculate $\epsilon_F$. The \texttt{find\char`_epsilon\char`_F.m} m-file uses the last energy level previously calculated by \texttt{pdeeig} as an initial guess to the built-in MATLAB root-finding \texttt{fzero} function. The \texttt{fzero} command solves the charge neutrality condition in \texttt{charge\char`_neutral.m}, which calculates the difference between the integrated electron density (Eq.~\eqref{E:elecNumDensity5}) and the integrated depletion region (cf. Fig.~\ref{F:facePinDepletionRegion}). If the $\epsilon_F$ value predicted by \texttt{find\char`_epsilon\char`_F.m} is higher than the last energy level previously calculated by \texttt{pdeeig}, the lower and upper limit of the \texttt{r} vector is incrementally changed and inputted to \texttt{pdeeig} until all the required wavefunctions/energies are obtained and Eq.~\eqref{E:chrgNeutral} is satisfied.

\subsection{Initial Poisson Equation and Self-Consistent Iteration}\label{ss:initPoissEq&iter}
With the wavefunctions and Fermi level computed, the summation $\sum_n |\psi_n(\tilde{x},\tilde{y})|^2 \sqrt{\epsilon_F-\epsilon_n}$ is computed by the \texttt{psi\char`_sqrt\char`_eps\char`_summation.m} m-file and stored in the global variable \texttt{psi\char`_sqrt\char`_eps\char`_sum} (which is used by the \texttt{n\char`_e\char`_func.m} m-file to compute $S_e$ in Eq.~\eqref{electron_donor_eV_simple}). It is worth mentioning that \texttt{n\char`_e\char`_func.m} utilizes the \texttt{griddata\char`_NaN.m} and \texttt{symmetrize\char`_coordinates.m} m-files to symmetrize the spatial electron density for the triangular and hexagonal geometries based on their respective two- and six-fold symmetries. The depletion region (cf. Fig.~\ref{F:facePinDepletionRegion}) is computed and stored in the global variable \texttt{heaviside\char`_n\char`_D}, which contains a vector having elements of 1 for $(\tilde{x},\tilde{y})$ values where $V_T(\tilde{x},\tilde{y}) \geq E_F$ and 0 otherwise. For hexagonal geometries that do not contain polarization effects, the $S_D(\tilde{x},\tilde{y})$ and $S_e(\tilde{x},\tilde{y})$ source terms are computed in the \texttt{n\char`_D\char`_func.m} and \texttt{n\char`_e\char`_func.m} m-files, respectively, and their sum is used as input for the ``\texttt{f}" coefficient in \texttt{assempde} to solve the dimensionless Poisson equation. For  both  of  the  Ga-  and  N-face  triangular  geometries, the sum of \texttt{n\char`_D\char`_func.m}, \texttt{n\char`_e\char`_func.m}, as well as the polarization source terms computed in \texttt{rho\char`_semipolar\char`_inside.m} and \texttt{rho\char`_strain\char`_analytic.m} are used as input for the ``\texttt{f}" coefficient in \texttt{assempde}. The inputs for the ``\texttt{c}" coefficient and boundary condition matrix, \texttt{b}, in \texttt{assempde} were discussed in Section~\ref{ss:initGuessPotential}.

With the initial potential computed by \texttt{assempde}, HADOKEN reinserts a fraction of the total potential (which includes an additional exchange--correlation term calculated by the \texttt{V\char`_xc.m} and \texttt{V\char`_total\char`_with\char`_xc.m} m-files) into a new Schr\"odinger equation to initialize the self-consistent procedure. To complete one cycle of the iteration scheme shown in Fig.~\ref{F:flowChart}, HADOKEN uses a 0.01 fraction of the potential for both the charge-neutral and Fermi-pinned hexagonal geometries, and a 0.05 fraction of the potential for triangular geometries. This cyclic process of simultaneously solving the Schr\"odinger and Poisson equations is continued until both are self-consistent, which we define as the situation where the average energy difference over all nodes of the electrostatic potential between successive iterations is less than $0.01\,\text{eV}$. To maintain a stable self-consistent feedback loop, the HADOKEN code uses an under-relaxation technique set by the variable \texttt{damping\char`_factor}, which is initially equal to 0.03 for hexagonal geometries and 0.05 for triangular geometries. As such, the inputted potential for the next iteration, \texttt{V\char`_poisson}, is calculated as \texttt{V\char`_poisson=V\char`_poisson\char`_old+damping\char`_factor*(V\char`_poisson-V\char`_poisson\char`_old)}, where \texttt{V\char`_poisson\char`_old} is the potential just computed. When the potential is nearly converged---defined to be the average energy difference of the potential being less than $0.01\mathcal{C}$---the \texttt{damping\char`_factor} variable is slowly increased to a maximum of 0.07 to further accelerate convergence. Once self-consistency is reached, the band-bending diagram, total electron density, and all occupied wavefunctions are output to the screen. Additionally, all variables used by HADOKEN are saved to a binary \texttt{.mat} file for further post-processing by the user. 

\section{Numerical Examples and Results}\label{results}
In the following subsections, we discuss typical calculations on a variety of nanowire geometries, configurations, and boundary conditions that can be performed with HADOKEN. Note that the parameters used as input to \texttt{set\char`_input\char`_parameters.m} are consistent with the reduced coordinates and scaling relations discussed in Section~\ref{theory}. Specifically, the first input variable, \texttt{vector\char`_of\char`_side\char`_lengths}, contains the side lengths of each interface expressed in units of $10\,\text{nm}$. The second input variable, \texttt{vector\char`_of\char`_V0}, contains the band edge energies in units of eV. The third input vector, \texttt{vector\char`_of\char`_masses}, contains the effective mass of each nanowire region in units of $m_0$, and the fourth input vector their static dielectrics. If a Fermi-pinning constraint is being enforced, the fifth input parameter, \texttt{epsilon\char`_F}, specifies the Fermi-level pinning in units of eV; otherwise, the last input required by \texttt{set\char`_input\char`_parameters.m} is the number of triangles used to discretize the NW cross-section in the finite element procedure. The doping density function used in the \texttt{set\char`_doping\char`_density.m} routine can accept any functional form and must be in units of ${10}^{18}\,{\text{cm}}^{-3}$. For instance, the user wanting to incorporate a doping density function that specifies which layers of the heterostructure are doped can use \texttt{heaviside\char`_core\char`_schrod.m} (see Section~\ref{methods}) as input to \texttt{set\char`_doping\char`_density.m}.

\subsection{Hexagonal Cross-Section} \label{ss:hexResults}
We first consider self-consistent Schr\"odinger--Poisson calculations for nonpolar core--shell NWs with hexagonal cross-sections.  As shown in Fig.~\ref{F:hexCross}, the spontaneous polarization contribution vanishes because the polarization axis is in the axial direction. The piezoelectric polarization also drops out---the strain components $\varepsilon_{xz}$ and $\varepsilon_{yz}$ are both zero since the displacements are uniform in the axial direction. It follows that all interfaces in the hexagonal NW are nonpolar, indicating that electron gas formation in these systems results exclusively from the conduction band edge discontinuity in $V_{\text{CB}}(\tilde{x},\tilde{y})$ and variations in the electrostatic potential, $V(\tilde{x},\tilde{y})$. Previous work by us and others have shown that electron gas formation is insensitive to local exchange--correlation effects in $V_{\text{xc}}(\tilde{x},\tilde{y})$ \cite{PAMELA,bJ02}.

\begin{figure}[H]
	\centering
	\begin{subfigure}[t]{\textwidth}
		\centering
		\includegraphics[width=\textwidth]{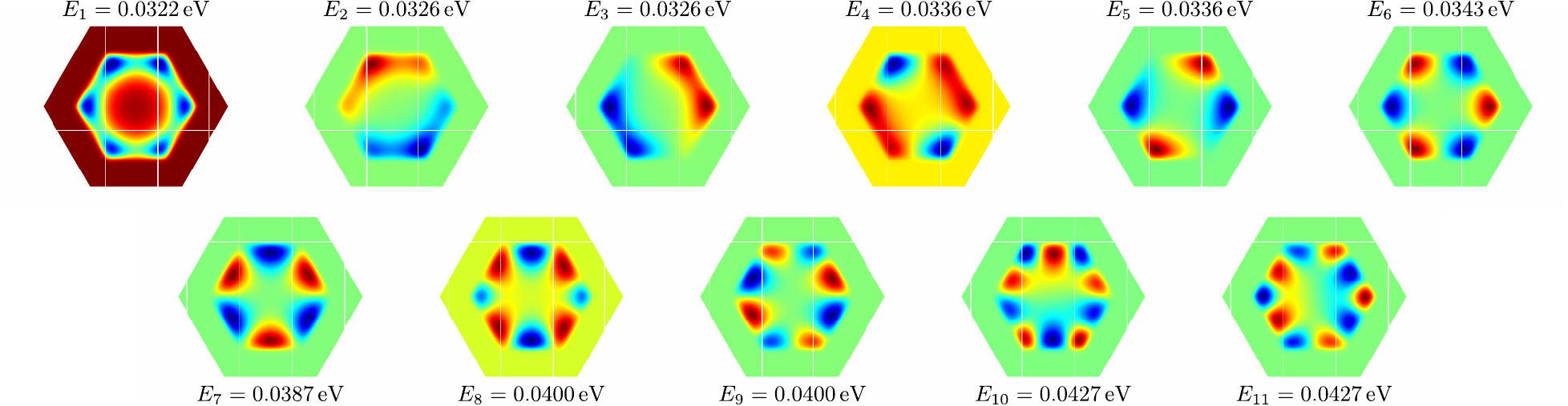}
		\caption{}
		\label{F:hexWF}
	\end{subfigure}
	\hfill
	\begin{subfigure}[t]{0.495\textwidth}
		\centering
		\includegraphics[width=66.146mm]{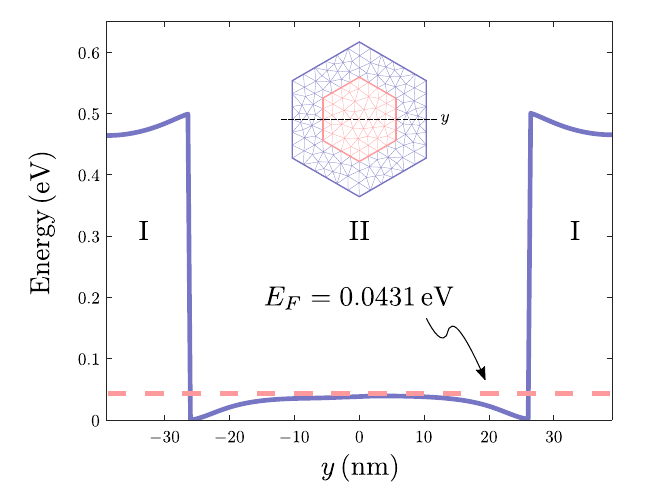}
		\caption{}
		\label{F:hexBandBend}
	\end{subfigure}
	\hfill
	\begin{subfigure}[t]{0.495\textwidth}
		\centering
		\includegraphics[width=66.146mm]{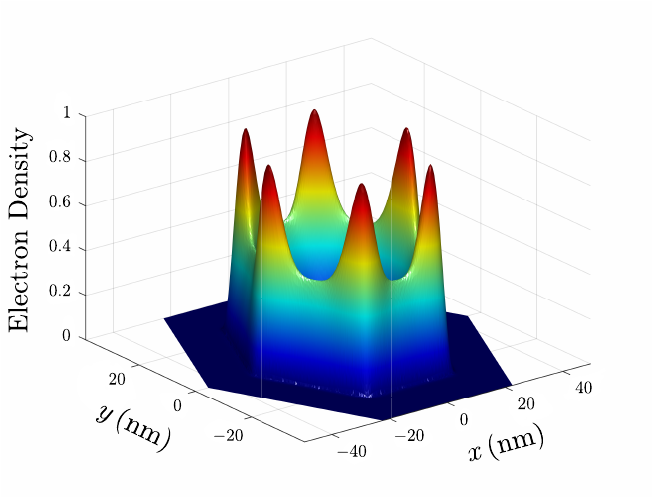}
		\caption{}
		\label{F:hexElecDensity}
	\end{subfigure}
	\caption{Calculated (a)\,wavefunctions, (b)\,band-bending diagram along the dashed line of the inset, and (c)\,charge distribution for a hexagonal cross section having a 30-nm core side length and 45-nm shell side length with a doping density of $\num{0.2e18} \, {\text{cm}}^{-3}$. The energies depicted in (a) are measured relative to the minimum of the conduction band, and the roman numerals in panel (b) indicate AlGaN (I) or GaN (II) regions along the $y$-axis.}
	\label{F:hex}
\end{figure}

The wavefunctions, band-bending diagram, and total electron densities shown in Fig.~\ref{F:hex} are outputted by the \texttt{main\char`_scf\char`_neumann.m} routine when the following parameters are used in the \texttt{set\char`_input\char`_parameters.m} and \texttt{set\char`_doping\char`_density.m} m-files, respectively:\\

\noindent
\begin{minipage}[t]{0.6\linewidth}
In \texttt{set\char`_input\char`_parameters.m}:
\vspace{-5pt}
\begin{verbatim}
vector_of_side_lengths=[4.5 3];
vector_of_V0=[0.5 0.0];
vector_of_masses=[0.2-0.12*0.3 0.2];
vector_of_eps=[9.28-0.61*0.3 9.28];
number_of_triangles=50000;
\end{verbatim}
\end{minipage} 
\hfill
\noindent
\begin{minipage}[t]{0.34\linewidth}
In \texttt{set\char`_doping\char`_density.m}:
\vspace{-5pt}
\begin{verbatim}
n_D=0.2;
\end{verbatim}
\end{minipage}\\
\\

\noindent
These input parameters correspond to a hexagonal NW with shell side length $s = 45 \, \text{nm}$, core side length $c = 30 \, \text{nm}$, and n-type doping density $n_D = \num{0.2e18} \, {\text{cm}}^{-3}$. For all results depicted in this and the following sections, 50,000 triangular elements were used to accurately capture the oscillating and highly localized wavefunctions at the core--shell interfaces.

It is worth noting that the spatial electron density shown in Figs.~\ref{F:hexWF} and \ref{F:hexElecDensity} is qualitatively different than the spatially uniform electron gas profile typically observed in macroscopic bulk/slab heterojunctions. Indeed, Fig.~\ref{F:hexElecDensity} shows the unique formation of six degenerate quasi-one-dimensional electron gases at vertices of the core--shell interface, which strongly resembles the lowest energy electron wavefunction. Specifically, the $E_1$ wavefunction corresponds to a highly localized charge distribution near the corners of the core--shell interface. As shown in Fig.~\ref{F:hexWF}, a few of the wavefunctions are doubly degenerate (i.e., $E_2$/$E_3$, $E_4$/$E_5$, $E_8$/$E_9$, and $E_{10}$/$E_{11}$), which arises from the irreducible representations of the $D_{6h}$ symmetry group (similar to that observed in benzene molecules). Furthermore, the other higher energy wavefunctions are also localized near the core--shell interface, such that their sum (cf. Eq.~\eqref{E:elecNumDensity5}) gives rise to a total electron distribution concentrated at the NW heterojunction's six corners. Note that HADOKEN can be used to explore other user-defined parameters; various combinations of core/shell sizes and doping densities can result in qualitatively distinct electron density profiles. Self-consistent calculations with low $n_D$ values, for example, tend to give relatively flat band-bending diagrams and a 2DEG localized in the core's \textit{center} (not shown in Fig.~\ref{F:hex}), rather than near the corners. Conversely, localization near the NW corners generally requires high values of $n_D$, particularly for small core sizes.

\begin{figure}[H]
	\centering
	\begin{subfigure}[t]{0.323\textwidth}
		\includegraphics[width=\textwidth]{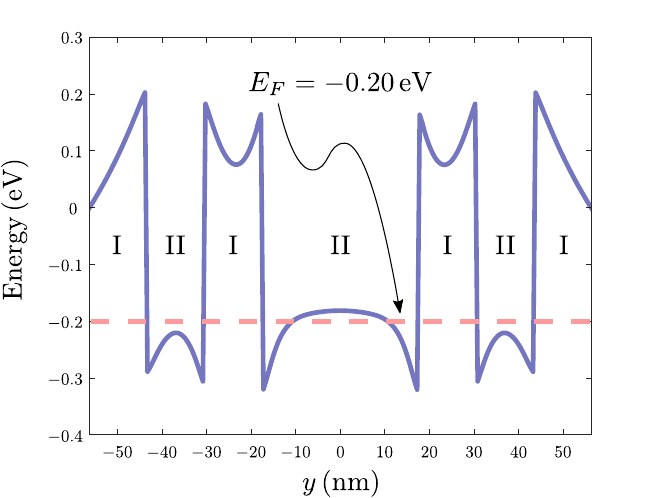}
		\caption{}
		\label{F:multiHexBandBend}
	\end{subfigure}
	\begin{subfigure}[t]{0.323\textwidth}
		\includegraphics[width=\textwidth]{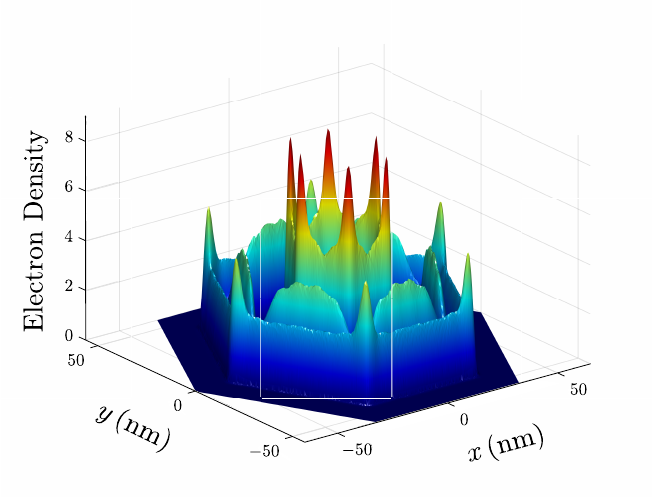}
		\caption{}
		\label{F:multiHexElecDensity3D}
	\end{subfigure}
	\begin{subfigure}[t]{0.323\textwidth}
		\includegraphics[width=\textwidth]{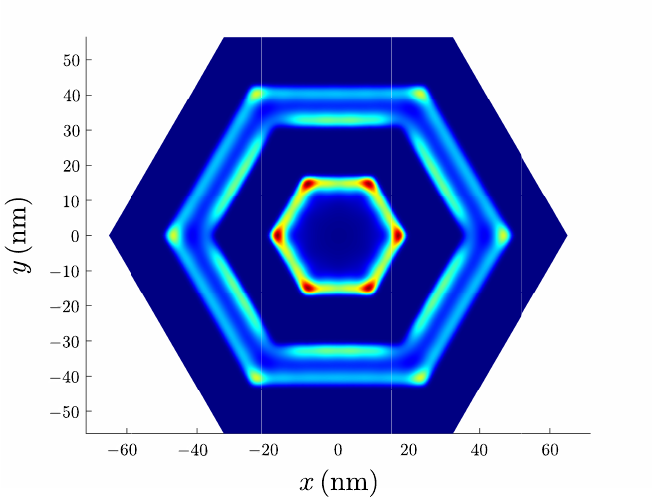}
		\caption{}
		\label{F:multiHexElecDensity2D}
	\end{subfigure}
	\caption{{(a)\,Band-bending diagram, (b)\,3D charge distribution, and (c)\,2D cross-sectional charge distribution for a core-multishell nanowire with a fixed Fermi level ($E_F=-0.2\,\text{eV}$) and doping density of $n_D(\tilde{x},\tilde{y}) = 5 \exp{[-0.1 (\tilde{x}^2+\tilde{y}^2)]}\,\text{cm}^{-3}$. The roman numerals in panel (a) delineate regions of AlGaN (I) from GaN (II) along the $y$-axis.}}
	\label{F:multiHex}
\end{figure}

To provide a more complex example of the various boundary conditions and geometries that HADOKEN can handle, Fig.~\ref{F:multiHex} depicts the self-consistent band-bending diagram and electron density for a core--multishell nanowire with a fixed Fermi level and more intricate doping density function.

The wavefunctions, band-bending diagram, and total electron densities shown in Fig.~\ref{F:multiHex} are output by running the \texttt{main\char`_scf\char`_dirichlet} command at the MATLAB prompt when the following input parameters are used in \texttt{set\char`_input\char`_parameters.m} and \texttt{set\char`_doping\char`_density.m} m-files, respectively:\\

\noindent
\begin{minipage}[t]{0.6\linewidth}
In \texttt{set\char`_input\char`_parameters.m}:
\vspace{-5pt}
\begin{verbatim}
vector_of_side_lengths=[6.5 5 3.5 2];
vector_of_V0=[0.5 0.0 0.5 0.0];
vector_of_masses=[0.2-0.12*0.3 0.2 0.2-0.12*0.3 0.2];
vector_of_eps=[9.28-0.61*0.3 9.28 9.28-0.61*0.3 9.28];
epsilon_F=-0.2;
number_of_triangles=50000;
\end{verbatim}
\end{minipage} 
\hfill
\noindent
\begin{minipage}[t]{0.34\linewidth}
In \texttt{set\char`_doping\char`_density.m}:
\vspace{-5pt}
\begin{verbatim}
n_D=5*exp(-0.1*(x.^2+y.^2));
\end{verbatim}
\end{minipage}\\
\\

\noindent
These input parameters correspond to a hexagonal nanowire with an n-type doping density of the form $n_D(\tilde{x},\tilde{y}) = 5 \exp{[-0.1 (\tilde{x}^2+\tilde{y}^2)]}\,\text{cm}^{-3}$, a fixed Fermi level of $-0.2\,\text{eV}$, an inner core/shell length of $20\,\text{nm}$/$35\,\text{nm}$, and an outer core/shell length of $50\,\text{nm}$/$65\,\text{nm}$. As shown in Figs.~\ref{F:multiHexElecDensity3D} and \ref{F:multiHexElecDensity2D}, the self-consistent total electron density in this core--multishell NW exhibits a much more complex structure. A total of twelve quasi-one-dimensional electron gases are visible, with the first six located at the inner quantum well's vertices and the other six at the outer quantum well's vertices. Moreover, a sheet-like distribution forms at each of the GaN/AlGaN interfaces of the outer quantum well.

\subsection{Triangular Cross-Sections}\label{ss:triResults}
\begin{figure}[H]
	\centering
	\begin{subfigure}[t]{\textwidth}
		\centering
		\includegraphics[width=\textwidth]{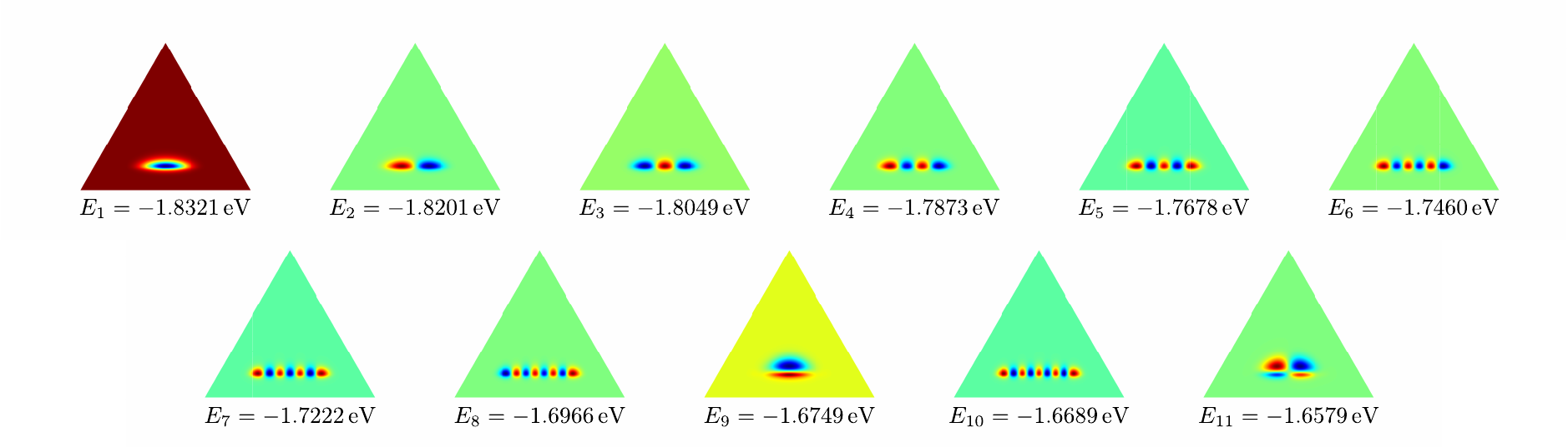}
		\caption{}
		\label{F:facePinWF}
	\end{subfigure}
	\hfill
	\begin{subfigure}[t]{0.49\textwidth}
		\centering
		\includegraphics[width=66.146mm]{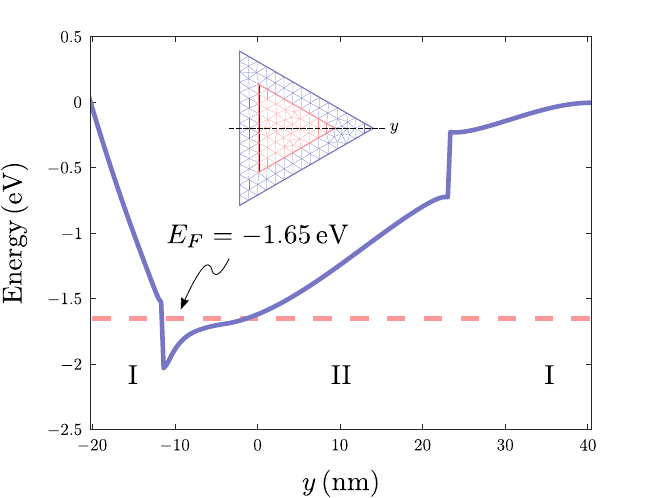}
		\caption{}
		\label{F:facePinBandBend}
	\end{subfigure}
	\hfill
	\begin{subfigure}[t]{0.49\textwidth}
		\centering
		\includegraphics[width=66.146mm]{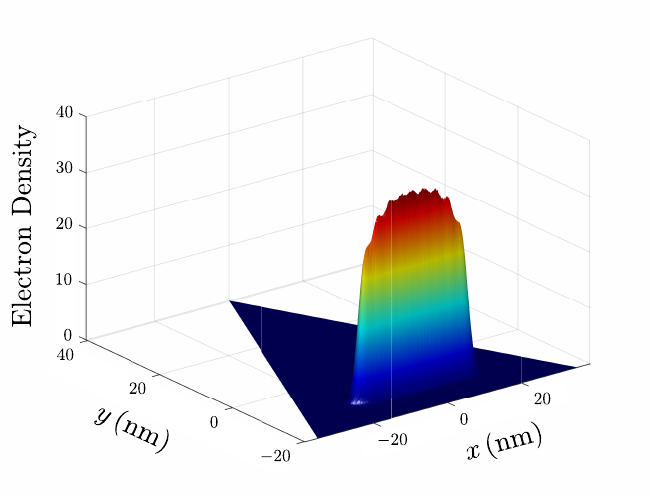}
		\caption{}
		\label{F:facePinElecDensity}
	\end{subfigure}
	\caption{Calculated (a)\,wavefunctions, (b)\,band-bending diagram along the dashed line of the inset, and (c)\,charge distribution for a $(0001)$ Ga-face triangular core--shell nanowire having a 40-nm core side length and 70-nm shell side length with a doping density of $\num{4.0e18} \, {\text{cm}}^{-3}$. The energies depicted in (a) are measured relative to the conduction band evaluated at the shell edge, and the roman numerals in panel (b) indicate AlGaN (I) or GaN (II) regions along the $y$-axis.}
	\label{F:facePin}
\end{figure}

Unlike the hexagonal cross-sections, core--shell NWs with triangular cross-sections possess both spontaneous and piezoelectric polarizations corresponding to the two orientations depicted in Fig.~\ref{F:triPolars}. The first case we discuss is the (0001) Ga-face triangular core--shell NW. Running the \texttt{main\char`_scf\char`_dirichlet} command at the MATLAB prompt in the \texttt{triangular\char`_Ga\char`_face\char`_coreshell} folder will output the wavefunctions and total electron densities shown in Fig.~\ref{F:facePin}. The input parameters to \texttt{set\char`_input\char`_parameters.m} and \texttt{set\char`_doping\char`_density.m} correspond to a triangular NW with n-type doping $n_D = \num{4e18} \, {\text{cm}}^{-3}$ and core and shell side lengths of $c = 40 \, \text{nm}$ and $s = 70 \, \text{nm}$, respectively.

For the Ga-face configuration, the spontaneous polarization induces a large positive surface charge at the $(0001)$ interface and smaller negative surface charges at the $(\bar{1}10\bar{1})$ and $(\bar{1}101)$ planes. This combination effectively attracts free electrons at the $(0001)$ interface, causing a 2DEG to accumulate at that GaN/AlGaN heterojunction, as shown in Fig.~\ref{F:facePinElecDensity}. Notice that the Ga-face triangular NW possesses a lower symmetry than the hexagonal NW discussed previously, and degenerate wavefunction pairs are not observed in this configuration. Instead, the lowest energy wavefunctions resemble one-dimensional particle-in-a-box-like patterns, with vertical nodal planes emerging as the energy of each wavefunction increases. At some critical energy (in this case, $E_9=-1.6749\,\text{eV}$), a horizontal nodal plane emerges, and the wavefunctions start to delocalize into other regions of the NW (see also $E_{11}$). Most notably, the sum of these wavefunctions via Eq.~\eqref{E:elecNumDensity5} gives rise to a sheet-like distribution of charge that is symmetric about the $x=0$ plane. We have used HADOKEN to explore other combinations of core/shell sizes and doping densities and found that a localized peak (rather than a delocalized sheet-like distribution) is obtained for small core sizes. For higher doping densities and larger core sizes, a sheet-like distribution emerges since the energy difference between occupied levels decreases for these scenarios, resulting in a semi-classical electron distribution resembling what is found in planar heterojunctions.

\begin{figure}[H]
	\centering
	\begin{subfigure}[t]{\textwidth}
		\centering
		\includegraphics[width=\textwidth]{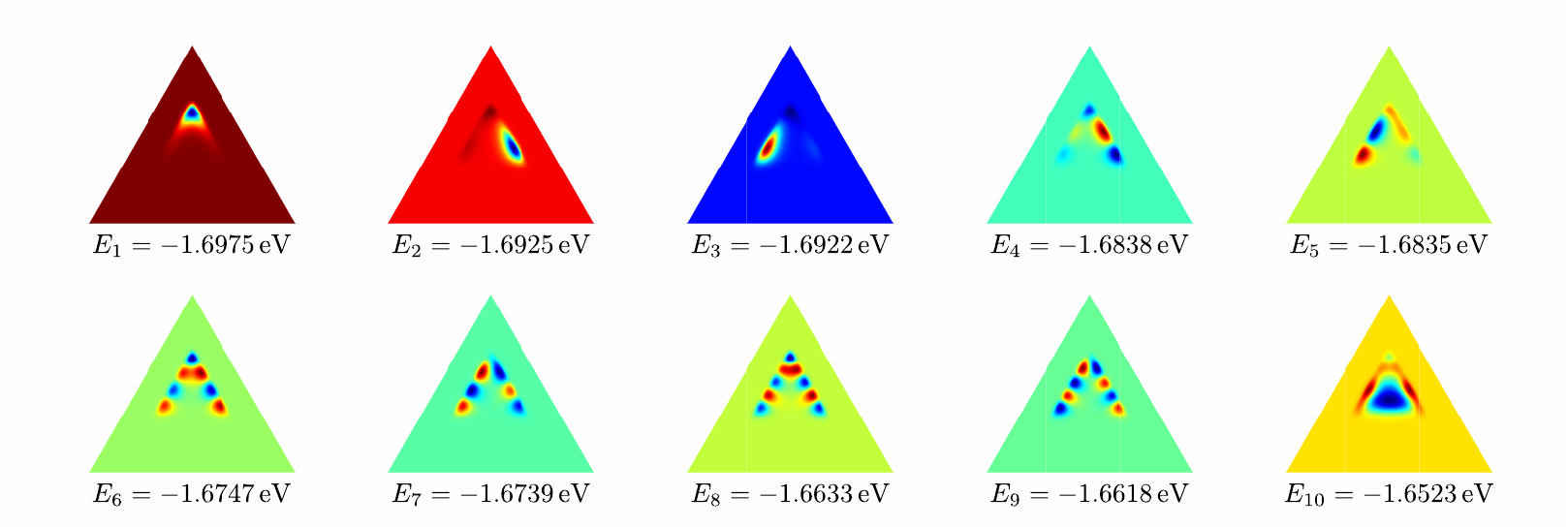}
		\caption{}
		\label{F:cornerPinWF}
	\end{subfigure}
	\hfill
	\begin{subfigure}[t]{0.495\textwidth}
		\centering
		\includegraphics[width=66.146mm]{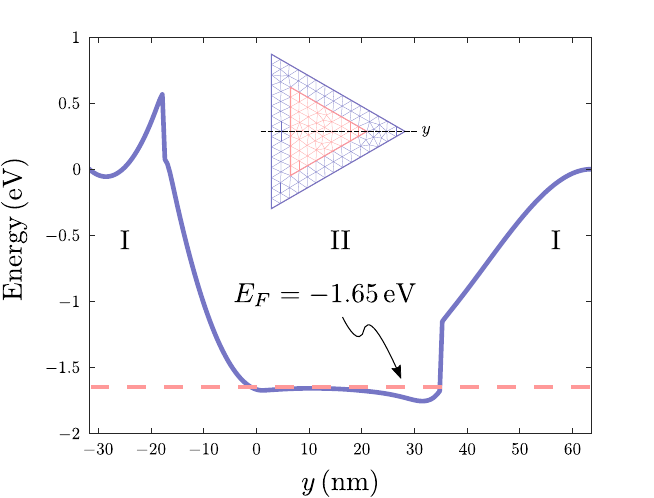}
		\caption{}
		\label{F:cornerPinBandBend}
	\end{subfigure}
	\hfill
	\begin{subfigure}[t]{0.495\textwidth}
		\centering
		\includegraphics[width=66.146mm]{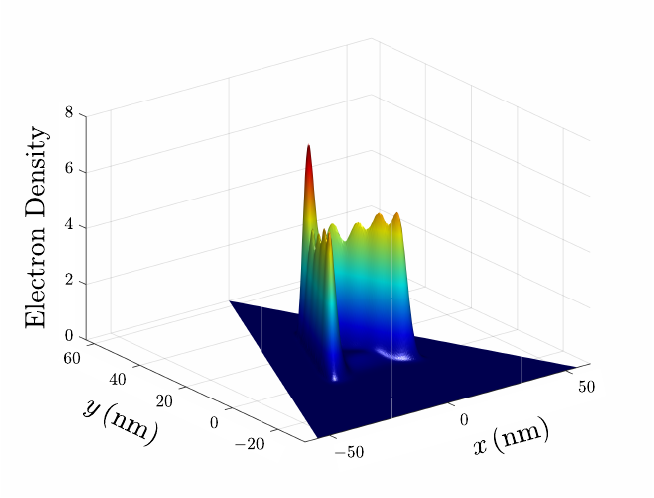}
		\caption{}
		\label{F:cornerPinElecDensity}
	\end{subfigure}
	\caption{Calculated (a)\,wavefunctions, (b)\,band-bending diagram along the dashed line of the inset, and (c)\,charge distribution for a $(000\bar{1})$ N-face triangular core--shell nanowire having a 60-nm core side length and 110-nm shell side length with a doping density of $\num{5.5e18} \, {\text{cm}}^{-3}$. The energies depicted in (a) are measured relative to the conduction band evaluated at the shell edge, and the roman numerals in panel (b) indicate AlGaN (I) or GaN (II) regions along the $y$-axis.}
	\label{F:cornerPin}
\end{figure}

The situation is materially different for the N-face orientation. Running the \texttt{main\char`_scf\char`_dirichlet} command at the MATLAB prompt in the \texttt{triangular\char`_N\char`_face\char`_coreshell} folder will output the wavefunctions and total electron densities shown in Fig.~\ref{F:cornerPin} when the following input parameters are used in \texttt{set\char`_input\char`_parameters.m} and \texttt{set\char`_doping\char`_density.m} m-files, respectively:\\

\noindent
\begin{minipage}[t]{0.6\linewidth}
In \texttt{set\char`_input\char`_parameters.m}:
\vspace{-5pt}
\begin{verbatim}
vector_of_side_lengths=[11 6];
vector_of_V0=[0.5 0.0];
vector_of_masses=[0.2-0.12*0.3 0.2];
vector_of_eps=[9.28-0.61*0.3 9.28];
epsilon_F=-1.65;
number_of_triangles=50000;
\end{verbatim}
\end{minipage} 
\hfill
\noindent
\begin{minipage}[t]{0.34\linewidth}
In \texttt{set\char`_doping\char`_density.m}:
\vspace{-5pt}
\begin{verbatim}
n_D=5.5;
\end{verbatim}
\end{minipage}\\
\\

\noindent
The input parameters to the \texttt{set\char`_input\char`_parameters.m} and \texttt{set\char`_doping\char`_density.m} routines correspond to a triangular NW having an n-type doping of $n_D = \num{5.5e18} \, {\text{cm}}^{-3}$ and core and shell side lengths of $c = 60 \, \text{nm}$ and $s = 110 \, \text{nm}$, respectively.

For the N-face configuration, the polarization results in a large negative surface charge to accumulate at the $(000\bar{1})$ interface and smaller positive surface charges along both semipolar interfaces. Consequently, donor electrons in the NW are repelled from the N-face but attracted to the adjacent positively-charged surfaces. The system reaches an electrostatic equilibrium by creating a localized electron gas near the vertex opposite to the negatively-charged $(000\bar{1})$ interface. It is also worth noting that a few of the lowest-energy wavefunctions shown in Fig.~\ref{F:cornerPinWF} are doubly (or almost doubly) degenerate. Specifically, the wavefunctions corresponding to $E_2$ and $E_3$ are nearly degenerate and resemble a geometric reflection of each other about the $x=0$ plane. The wavefunctions corresponding to $E_4$ and $E_5$ are also nearly degenerate, with the former having two horizontal nodal planes and the latter having one horizontal nodal plane and one vertical nodal plane at $x=0$. Similarly, the wavefunctions corresponding to $E_6$ and $E_7$ are almost nearly degenerate, with the former having three horizontal nodal planes and the latter having two horizontal nodal planes and one vertical nodal plane at $x=0$. As the energy of the individual wavefunctions increases, the charge distribution expands further down each semipolar face, leading to a total electron density that is primarily localized at the vertex of the triangular NW. Additional calculations with the HADOKEN code (not shown in Fig.~\ref{F:cornerPin}) have shown that small core sizes tend to favor quantum 1DEGs, since the magnitude of the electron density at the vertex decreases (and starts to extend symmetrically along the two semipolar faces) as the core size increases.

\section{Conclusions}\label{conclusions}
In this contribution, we have provided and extensively documented an open-source software code for predicting two-dimensional electron gas formation in heterostructure core--shell nanowires. The algorithms in the HADOKEN software utilize a robust finite element procedure that solves coupled Schr\"odinger and Poisson equations self-consistently for a variety of geometries, doping densities, and external boundary conditions. Most importantly, the HADOKEN software can be downloaded from the Computer Physics Communications International Computer Program Library to investigate material composition effects, bandgap alignment, doping density, and cross-sectional size on Fermi gas formation in a variety of nanowire configurations. In addition, the user-friendly MATLAB code serves as a starting point for researchers that may need minor modifications of the well-documented source code to simulate other materials and geometries beyond those discussed in this work.

Looking forward, we anticipate that the HADOKEN software package could be used in a variety of other applications that require electronic structure calculations of these unique structures. For example, since our calculations demonstrate that electron gases at nanoscale core--shell interfaces differ significantly from their bulk counterparts, we anticipate that other observables such as electron transport \cite{doi:10.1021/acsami.9b01400} or optical properties \cite{doi:10.1021/acs.jctc.7b00423} in these systems would also exhibit unique behavior. As such, the wavefunctions and total electron densities computed by the HADOKEN code can serve as a starting point to initialize the computation of these dynamical properties. Similarly, the self-consistent algorithms in the HADOKEN code can also be further parallelized or modified to include other many-body effects (such as nonlocal exchange--correlation effects \cite{PAMELA, bW10}) that may have a significant influence on electron localization effects observed in these systems. The open-source HADOKEN software code enables a path forward to explore these other properties as well as provides an easy-to-use, predictive tool to understand and modulate electron confinement effects in these unique nanosystems.

\section{Acknowledgements}\label{acknowledgements}
C. C. acknowledges support from the National Science Foundation INTERN program under Grant No. CHE-2028365.


\setcounter{section}{0}
\setcounter{equation}{0}
\renewcommand{\theequation}{\thesection.\arabic{equation}}

\bibliographystyle{elsarticle-num}
\bibliography{mainb}

\end{document}